# Quantitative determination of optical and recombination losses in thin-film photovoltaic devices based on external quantum efficiency analysis

Akihiro Nakane,[1] Hitoshi Tampo,[2] Masato Tamakoshi,[1] Shohei Fujimoto,[1] Kang Min Kim,[2] Shinho Kim,[2] Hajime Shibata,[2] Shigeru Niki,[2] and Hiroyuki Fujiwara[1,a]

[1]Department of Electrical, Electronic and Computer Engineering, Gifu University, 1-1 Yanagido, Gifu 501-1193, Japan
[2]Research Center for Photovoltaics, National Institute of Advanced Industrial Science and Technology (AIST), Central2, 1-1-1 Umezono, Tsukuba, Ibaraki 305-8568, Japan

**Abstract**
In developing photovoltaic devices with high efficiencies, quantitative determination of the carrier loss is crucial. In conventional solar-cell characterization techniques, however, photocurrent reduction originating from parasitic light absorption and carrier recombination within the light absorber cannot be assessed easily. Here, we develop a general analysis scheme in which the optical and recombination losses in submicron-textured solar cells are evaluated systematically from external quantum efficiency (EQE) spectra. In this method, the optical absorption in solar cells is first deduced by imposing the anti-reflection condition in the calculation of the absorptance spectrum, and the carrier extraction from the light absorber layer is then modeled by considering a carrier collection length from the absorber interface. Our analysis method is appropriate for a wide variety of photovoltaic devices, including kesterite solar cells [$Cu_2ZnSnSe_4$, $Cu_2ZnSnS_4$, and $Cu_2ZnSn(S,Se)_4$], zincblende CdTe solar cells, and hybrid perovskite ($CH_3NH_3PbI_3$) solar cells, and provides excellent fitting to numerous EQE spectra reported earlier. Based on the results obtained from our EQE analyses, we discuss the effects of parasitic absorption and carrier recombination in different types of solar cells.

[a]Author to whom correspondence should be addressed. Electronic mail: fujiwara@gifu-u.ac.jp.

## I. INTRODUCTION

Conversion efficiency of solar cells is influenced by a variety of physical and optical properties of constituent layers and, in solar cell optimization, it is necessary to identify and improve the limiting factors. In photovoltaic devices, the extraction of photo-generated carriers from a semiconductor light absorber is an essential process and the presence of recombination centers within absorber layers deteriorates the short-circuit current density ($J_{sc}$) significantly.[1] The effect of the recombination generally becomes more severe in the solar-cell bottom region, where the generated carriers need to reach the front interface through carrier diffusion. In other words, solar cell efficiencies are often limited by the diffusion length ($L_D$) of photo-generated carriers.[1]

Unfortunately, the evaluation of $L_D$ in thin-film solar cells has been rather challenging. Although the $L_D$ of solar cells can be evaluated from electron-beam induced current (EBIC)[2-5] and photoluminescence (PL) lifetime[6-8] measurements, the interpretation of the EBIC and PL data is not straight forward. Specifically, in the EBIC characterization of thin-film solar cells, the electron beam size is often comparable to the absorber thickness, and the EBIC signal may change by the effects of surface recombination and injection density.[9] The PL signals are also influenced by multiple factors, including bulk life time, interface recombination, carrier mobility, and injection/doping density.[10] Moreover, for CuInGaSe$_2$ (CIGSe) solar cells, the PL lifetime decreases by two orders of magnitude after the CIGSe surface is exposed to air for 1 day and the strong laser-light illumination induces the degradation.[11]

One alternative method that can be used for the accurate $L_D$ evaluation is the quantum efficiency (QE) analysis.[12-18] Since the QE spectra contain the depth information due to the wavelength ($\lambda$)-dependent penetration depth of light, the $L_D$ can be deduced from the QE response in the long $\lambda$ region near the band gap ($E_g$).[15-18] By analyzing the QE spectra obtained at different bias voltages, the recombination mechanism can further be studied.[18] In the conventional QE characterization, however, the QE spectra are analyzed without considering optical effects induced by light scattering and parasitic light absorption in the transparent conductive oxide (TCO) and metal back electrode. Thus, the optical losses within solar cell devices are not generally clear.

In our previous study, on the other hand, we have developed an external QE (EQE) simulation method that can be applied to the explicit optical simulation of submicron-textured CIGSe solar cells.[19] Although the EQE simulation generally becomes difficult when textured structures are present,[20-22] our method allows the accurate determination of the EQE spectra without considering three-dimensional structures. This method is based on the finding that the optical effect of submicron rough textures is mainly the elimination of interference pattern.[19] In the simulation of this method, the optical response of the light absorber is estimated by simply forcing the anti-reflection condition in a perfectly flat optical system. From this simple optical simulation method using the anti-reflection condition (hereafter denoted as ARC method), the EQE spectrum of CIGSe solar cells fabricated by a standard three-stage evaporation process has been reproduced almost perfectly.[19]

Quite fortunately, the ARC method is a general method and can be applied to other thin-film photovoltaic devices without any restrictions. Since accurate optical simulation for arbitrary thin film structures can now be performed based on the ARC method, if the carrier extraction or recombination is further assumed in this simulation, both optical and recombination losses can be evaluated simultaneously from conventional EQE spectra. Such EQE analyses can be employed systematically to fill the strong need for revealing limiting physical/optical factors in solar cell devices.

In this study, we have established a general EQE analysis method in which the effect of the carrier recombination in the light absorber layer is further modeled within the framework of the ARC method. In this extended ARC method (hereafter denoted as e-ARC method), the carrier extraction from the light absorber is expressed by a simple exponential decaying function. Although the determination of $L_D$ is preferred, we have focused on the quantitative analysis of the recombination loss, and we have greatly simplified the QE analysis by considering only the effective carrier-collection length from the absorber front interface. Even though the recombination mechanism cannot be characterized from our EQE analysis, the developed e-ARC method is quite effective in determining the recombination loss that occurs in solar-cell bottom region. We demonstrate that the e-ARC method can be employed to deduce $J_{sc}$ losses induced by parasitic optical absorption and carrier recombination in a variety of solar cell devices including $Cu_2ZnSnSe_4$ (CZTSe),[16,23-30] $Cu_2ZnSnS_4$ (CZTS),[31-39] CZTSSe,[17,40-47] CdTe,[48] and $CH_3NH_3PbI_3$ hybrid perovskite[49,50] solar cells.

**II. EXPERIMENT**

For the EQE analysis of a CZTSe solar cell, we fabricated a device consisting of (Al grid)/ZnO:Al/non-doped ZnO/CdS/CZTSe/Mo/soda lime glass substrate.[51] The CZTSe layer in the solar cell was prepared by a single coevaporation process of Cu, Zn, Sn and Se elementary sources at 340 °C, followed by thermal annealing at 550 °C under Se and SnSe$_2$ atmosphere. We determined the composition of the CZTSe layer from electron probe microanalyzer (Cu:Zn:Sn:Se = 1.80:1.31:0.89:4.00) and, for this layer, the compositional ratios of Cu/(Zn+Sn) = 0.82 and Zn/Sn = 1.47 were estimated. In the fabrication of the CZTSe solar cell, the CdS layer was formed on the CZTSe layer by a standard chemical-bath deposition technique,[52] whereas we fabricated the non-doped ZnO, ZnO:Al and Mo layers by sputtering.

Figure 1 shows the cross-sectional transmission electron microscopy (TEM) images of the CZTSe solar cell fabricated in this study: (a) the whole solar-cell structure, (b) the enlarged image for the ZnO/CdS/CZTSe interface, and (c) the enlarged image for the CZTSe/Mo interface. When the CZTSe layer is prepared by the single coevaporation, the void-rich interface structure is formed on the substrate. Similar non-uniform interface structures have been observed in other CZTSe solar cells.[28,30,53] The CZTSe solar-cell structure similar to Fig. 1(a) has also been confirmed in the corresponding scanning electron microscope (SEM) images. However, the void structure at the interface is more pronounced in the TEM images, as the focused-ion-beam process, used in the TEM sample preparation, tends to increase the interface void fraction.

In the TEM image of Fig. 1(b), the polycrystalline CdS layer covers the CZTSe layer uniformly with a thickness of 70 nm. Moreover, a uniform layer with a thickness of 150 nm is present between the CZTSe and Mo layers [Fig. 1(c)]. From the compositional analysis of this layer, we confirmed that this layer is MoSe$_2$, in agreement with earlier CZT(S)Se studies.[17,25,53-55] In the initial deposition process, therefore, the MoSe$_2$ layer is first formed on the Mo substrate by the excessive Se supply,[55] followed by the CZTSe formation on the MoSe$_2$ layer. The generation of the void-rich structure at the interface could be related to the poor adhesion of the CZTSe to the MoSe$_2$ layer. It should be noted that the thickness of the MoSe$_2$ layer formed during the CZTSe growth is much thicker than that formed during the CIGSe growth.[19]

For the CZTSe solar cell of Fig. 1, we obtained a conversion efficiency of 8.46% with a $J_{sc}$ of 33.4 mA/cm$^2$, an open-circuit voltage ($V_{oc}$) of 418 mV and a fill factor (*FF*) of 0.606. In the EQE measurement of this solar cell, however, we obtained a slightly smaller $J_{sc}$ value of 31.6 mA/cm$^2$ in part due to the difference in the light illumination conditions between the current-voltage and EQE measurements. Specifically, in the EQE measurement, the effect of the carrier recombination is expected to become more significant since the photocurrent generated at a selected $\lambda$ is much weaker than that under AM1.5G illumination. In this study, we have analyzed the EQE spectrum obtained from the CZTSe solar cell in Fig. 1 to establish a general EQE analysis procedure.

## III. EQE ANALYSIS

**A. Calculation method**

The EQE spectra of all solar cells, including CZTSe, CZTS, CZTSSe, CdTe and $CH_3NH_3PbI_3$ solar cells, were calculated based on the e-ARC method established in this study. In this method, the optical response of solar cell structures is calculated first from the conventional optical-admittance method[56-58] assuming a perfectly flat device structure. Figure 2(a) indicates the calculation procedure of this optical admittance method. In this figure, $N_j$ shows the complex refractive index ($N_j = n_j - ik_j$) defined from the refractive index $n$ and the extinction coefficient $k$, whereas $Y_j$, $\psi_j$, and $d_j$ denote the optical admittance, potential transmittance, and thickness of the $j$th layer, respectively. In this optical model, $j = 0$ corresponds to the ambient environment (air).

Using the magnetic field ($H_f$) and electric field ($E_f$), the optical admittance is expressed as $Y = H_f/E_f$. For non-absorbing media ($k = 0$), there is a general relation of $H_f = nE_f$ and the same relation holds for $N$ ($H_f = NE_f$). Thus, $Y$ basically represents $N$, and $Y$ of the substrate is expressed as $Y_j = N_{j+1}$. From a matrix calculation that satisfies the boundary conditions of ($E_f$, $H_f$), $Y_j$ at the interface of $N_j/N_{j+1}$ can be transferred to $Y_{j-1}$ by

$$Y_{j-1} = \frac{Y_j \cos\delta_j + iN_j \sin\delta_j}{\cos\delta_j + iY_j \sin\delta_j / N_j}. \tag{1}$$

Here, $\delta_j$ denotes the phase thickness expressed by $\delta_j = 2\pi N_j d_j/\lambda$. If the similar calculation is repeated in a multilayer stack, we finally obtain $Y_0$, from which the reflectance of the flat structure ($R_{\text{flat}}$) is calculated as

$$R_{\text{flat}} = |1 - Y_0|^2 / |1 + Y_0|^2. \tag{2}$$

From Eqs. (1) and (2), therefore, $R_{\text{flat}}$ can be obtained relatively easily if $N_j$ and $d_j$ of all the optical layers are known.

In the optical admittance method, the light absorption in each layer is determined by considering the light transmittance ($T$) at each interface. If the light intensity is $I = 1$, $T$ at the top surface is expressed by $T = 1 - R_{\text{flat}}$. The $T$ value at the interface of the second or higher layer can then be calculated by multiplying $\psi$ of each layer sequentially from the top layer, as shown in Fig. 2(a), and $\psi_j$ is given by

$$\psi_j = \frac{\text{Re}(Y_j)}{\text{Re}(Y_{j-1})|\cos\delta_j + iY_j \sin\delta_j / N_j|^2}. \tag{3}$$

From $R_\text{flat}$ and $\psi$ of each layer, the absorptance in the $j$th layer ($A_j$) is expressed by

$$A_j = (1-R_\text{flat})(1-\psi_j)\prod_{k=1}^{j-1}\psi_k. \quad (4)$$

Based on the above procedure, the absorptance in each solar-cell component layer is determined from the top to bottom layer ($A_1 \rightarrow A_j$).

Figure 2(b) shows the optical model constructed for the CZTSe solar cell in Fig. 1. In this model, the layer thicknesses deduced from the TEM measurements in Fig. 1 are shown. However, the thickness of the non-doped ZnO layer cannot be distinguished in the TEM image and was determined from a thickness of a single ZnO layer formed separately on a different substrate (50 nm). The optical effect of the void-rich rear interface structure was expressed by assuming a void volume fraction of 30 vol.% in the CZTSe bottom layer with a thickness of 310 nm. To calculate the optical properties of this bottom layer, we employed the Bruggeman effective-medium approximation.[59,60] From the EQE simulations, we confirmed that the void volume fraction in the bottom region does not affect the EQE spectrum of the CZTSe solar cell significantly.

Figure 3 summarizes the calculation procedure of the e-ARC method. This EQE simulation was performed by the optical model of Fig. 2(b) using reported optical constants of CZTSe,[61] ZnO:Al,[19] non-doped ZnO,[19] CdS,[19,62] MoSe$_2$,[19,63] and Mo.[19] In Fig. 3(a), $R_\text{flat}$ of the CZTSe solar cell, calculated from Eqs. (1) and (2), is shown. When the flat layer structure is assumed, the absolute value of $R_\text{flat}$ varies largely versus $\lambda$ by the optical interference effect. In Fig. 3(a), $A_\text{flat}$ represents the $A$ spectrum of the CZTSe layer obtained from $R_\text{flat}$ using Eq. (4). In this case, $A_\text{flat}$ is modulated strongly by the large optical interference, and the absolute values of $A_\text{flat}$ reduce in the $\lambda$ region where $R_\text{flat}$ is large. It should be emphasized that the $A_\text{flat}$ spectrum obtained assuming the flat structure is quite different from the EQE spectra reported for CZTSe solar cells.[16,23-30] The large discrepancy can be attributed to the presence of natural textures (see Fig. 1), which induces relatively large light scattering.

In our previous study for CIGSe solar cells, we found that the optical effect of the submicron texture is mainly the elimination of the optical interference pattern and can be approximated by linearly connecting the minimum positions that appear in $R_\text{flat}$.[19] In Fig. 3(a), $R_\text{tex}$ represents the reflectance spectrum obtained from this procedure and the minimum $R_\text{flat}$ positions are denoted by the red circles. By simply connecting these red circles, the $R$ spectrum for a submicron-textured structure ($R_\text{tex}$) can be deduced without any knowledge of geometrical structures. The $A$ spectrum corresponding to the textured structure ($A_\text{tex}$) can then be obtained easily by replacing $R_\text{flat}$ in Eq. (4) with $R_\text{tex}$. In the optical model of Fig. 2(b), however, there is the CZTSe/void layer, in addition to the uniform CZTSe layer. For this structure, $A_{\text{flat}}$ and $A_{\text{tex}}$ in Fig. 3(a) have been obtained by simply adding the corresponding $A$ spectra obtained from these two layers.

In a multilayer system, the minimum positions in the $R_{\text{flat}}$ spectrum basically satisfy the anti-reflection condition, in which the multiply reflected beams generated within a layered structure are out of phase. In the established method, the anti-reflection condition is superimposed on the result of $R_{\text{flat}}$ to express the interference-fringe elimination by natural textures. It can be seen from Fig. 3(a) that the yellow area, which corresponds to the difference between $A_{\text{flat}}$ and $A_{\text{tex}}$, is absorbed additionally by assuming the anti-reflection condition. From the above approach (ARC method), the EQE spectra of CIGSe solar cells are reproduced almost perfectly.[19] The calculation of textured solar cells can be performed more easily by using experimental $R$ spectra obtained from actual solar cell structures. In this case, the experimental $R$ spectrum is used directly in the calculation of Eq. (4), instead of $R_{\text{tex}}$.[19]

In this study, by further modeling the carrier collection within the light absorber layer, we have developed the e-ARC method. To simplify the analysis, we describe the carrier collection efficiency ($H$) by considering the carrier collection length ($L_C$) from the absorber interface as follows:

$$H_1(\lambda) = 1 - \exp[-\alpha(\lambda)L_C]. \tag{5}$$

Here, $H_1(\lambda)$ denotes the carrier collection model expressed by Eq. (5) and the $\alpha$ shows the absorption coefficient of absorber layers. This model is equivalent to the light absorption expressed by the Beer's law, $A(\lambda) = 1 - \exp[-\alpha(\lambda)d]$, where $d$ shows the distance from the surface (or interface).[59] The $H_1(\lambda)$ obtained from Eq. (5) represents the light absorption at depth $L_C$ from the surface of materials having infinite thickness. In this model, therefore, the optical confinement effect, induced by the back-side reflection and the resulting multiple light reflection within the absorber layer, is neglected completely. Using Eq. (5), the simple optical simulation of internal QE (IQE) was also performed,[1,64] but such efforts have been limited. In Eq. (5), $L_C$ can be considered as an effective length where minority carriers in the absorber layer are extracted efficiently and the relation between $L_C$ and $L_D$ is expressed as

$$L_C = L_D + W, \tag{6}$$

where $W$ indicates the depletion layer thickness.[1] In the model of $H_1(\lambda)$, the influences of the carrier diffusion and carrier drift in the space-charge region are not described separately. Moreover, the effects of the band offset and the carrier recombinations that occur in the bulk, interface (front or rear) and grain-boundary regions are not specifically modeled. The significance of the $H_1(\lambda)$ model is that all the complex effects concerning the carrier collection are represented by a single analysis parameter of $L_C$.

Figure 3(b) shows $H_1(\lambda)$ calculated from $\alpha(\lambda)$ of the CZTSe using Eq. (5). When $L_C$ is infinite ($L_C = \infty$), we obtain $H_1(\lambda) = 1$ in the energy region above the absorption edge of the CZTSe [$\lambda \leq 1770$ nm in Fig. 3(b)]. As $L_C$ decreases, the $H_1$ value in the longer $\lambda$ region gradually reduces. In the calculation of $H_1(\lambda)$ using Eq. (5), however, it is necessary to use unrealistic $L_C$ values that are far larger than the total thickness of the absorber layer (~1 $\mu$m in Fig. 1). In particular, to obtain the sufficiently high $H_1(\lambda)$ values in the low $\alpha$ region, quite high $L_C$ values need to be used. Moreover, when the optical confinement is strong, $L_C$ values obtained by applying Eq. (5) also increase, as the effective optical-pass length increases in this case. Thus, $L_C$ defined by Eq. (5) should be considered as a parameter (reference) value. However, when $L_C$ is smaller than the absorber layer thickness, $L_C$ approximates a real physical value.

Although $H_1(\lambda)$ in Eq. (5) is a rather practical model, a more exact equation for $H(\lambda)$ has been derived by solving the carrier continuity equation:[12,13]

$$H_2(\lambda) = 1 - \exp[-\alpha(\lambda)W]/[1 + \alpha(\lambda)L_D]. \tag{7}$$

In this study, the carrier collection model expressed by Eq. (7) is denoted as $H_2(\lambda)$. If this model is applied for the EQE analysis, the parameters ($W$, $L_D$) can be deduced from the fitting analysis of the EQE spectrum. Nevertheless, Eq. (7) has also been derived by neglecting the optical interference effect in absorber layers. So far, using $H_2(\lambda)$, simple QE analyses have been made assuming $H_2(\lambda) = IQE(\lambda)$ [or $EQE(\lambda)$].[15-17] It should be noted that, if $L_D = 0$ is assumed in the models of $H_1(\lambda)$ and $H_2(\lambda)$ [i.e., $L_C = W$ in Eqs. (5) and (6)], both models are reduced to the same form of $H(\lambda) = 1 - \exp[-\alpha(\lambda)W]$. In the model of $H_1(\lambda)$, however, the carrier collection by the carrier drift and diffusion is approximated by using a single parameter of $L_C$ and the relation of $L_C = W$ is not always valid.

In this study, the EQE analyses are performed by employing both $H_1(\lambda)$ and $H_2(\lambda)$. However, the analysis and interpretation of the data are simplified greatly when $H_1(\lambda)$ is used. Thus, we implemented the EQE analyses of various solar cells by using $H_1(\lambda)$, unless otherwise stated. The EQE analysis using $H_1(\lambda)$ is also advantageous when the two parameters ($W$, $L_D$) in $H_2(\lambda)$ show high correlation in the EQE fitting analysis.

In the e-ARC method, the EQE spectrum is estimated quite simply as

$$EQE(\lambda) = A_{tex}(\lambda)H(\lambda). \tag{8}$$

Figure 3(c) shows the EQE spectra calculated from the e-ARC method using $A_{tex}$ in Fig. 3(a) and $H_1(\lambda)$ in Fig. 3(b). In the case of $L_C = \infty$ [or $H_1(\lambda) = 1$], we obtain $EQE(\lambda) = A_{tex}(\lambda)$. This EQE spectrum corresponds to the one calculated from the ARC method, in which no carrier recombination is assumed to occur. In the EQE spectra calculated from the e-ARC method, however, the EQE response in the longer $\lambda$ region decreases as the $L_\mathrm{C}$ value reduces due to the limited carrier extraction. From the EQE spectrum calculated using the e-ARC method, the recombination loss can be estimated further as a difference from a baseline EQE spectrum of $L_\mathrm{C} = \infty$. In the developed method, therefore, the recombination loss is calculated quantitatively using $L_\mathrm{C}$ as a sole parameter. It should be noted that, when the total thickness of the light absorber is modified, the optical interference and back-side reflection in solar cells change significantly.[65] Thus, it is necessary to model the carrier generation and collection separately.

From the EQE spectrum deduced by Eq. (8), $J_\mathrm{sc}$ of the solar cell is finally calculated as

$$J_{sc} = \frac{e\lambda}{2\pi\hbar c}\int EQE(\lambda)F(\lambda)d\lambda, \qquad (9)$$

where $e$ and $c$ are the electron charge and the speed of light, respectively. In Eq. (9), $F(\lambda)$ indicates the solar irradiance in units of $\mathrm{W\,cm^{-2}\,nm^{-1}}$ under AM1.5G illumination.

**B. Dielectric function**

Figure 4 summarizes (a) the $\varepsilon_2$ spectra of the dielectric functions ($\varepsilon = \varepsilon_1 - i\varepsilon_2 = N^2$) and (b) the $\alpha$ spectra of $\mathrm{Cu_2SnSe_3}$ (CTSe), CZTSe, $\mathrm{CuInSe_2}$ (CISe), CZTS, $\mathrm{CuGaSe_2}$ (CGSe), CdTe and $\mathrm{CH_3NH_3PbI_3}$ absorber layers used for the EQE analyses and simulations. These dielectric functions are the reported data of CTSe,[61] CZTSe,[61] CISe,[66,67] CZTS,[68] CGSe,[66,67] CdTe,[69] and $\mathrm{CH_3NH_3PbI_3}$.[65] The onset of the light absorption ($\varepsilon_2 > 0$) basically corresponds to $E_\mathrm{g}$ and, for the Cu-Se semiconductors, the $E_\mathrm{g}$ values are 0.68 eV (CTSe),[61] 0.91 eV (CZTSe),[61] 1.00 eV (CISe),[66] 1.32 eV (CZTS),[68] and 1.70 eV (CGSe).[66] It can be seen that the $\varepsilon_2$ spectral shapes of the CTSe, CZTSe and CZTS are quite similar. The $E_\mathrm{g}$ values of CdTe and $\mathrm{CH_3NH_3PbI_3}$ are reported to be 1.49 eV (Ref. 70) and 1.61 eV,[65] respectively.

Quite interestingly, all the light absorbers in Fig. 4(b) show similar $\alpha$ values near $E_\mathrm{g}$ ($\alpha \sim 1 \times 10^4\,\mathrm{cm^{-1}}$), although the CZTSe and CTSe semiconductors indicate large $\alpha$ values exceeding $10^5\,\mathrm{cm^{-1}}$ at 2.0 eV. It can be seen that CdTe and $\mathrm{CH_3NH_3PbI_3}$ absorbers exhibit quite sharp absorption tail, compared with the kesterite semiconductors. The density functional theory (DFT) analysis of CZTSe and CZTS shows that intermixing of constituent atoms occurs rather easily for the Cu, Zn and Sn atoms,[71,72] which may contribute to generate the tail states in these semiconductors.

The earlier DFT studies reveal that the valence bands of the CZTSe (CZTS) are the anti-bonding states formed by the $p$–$d$ hybridization of the Cu $3d$ and Se $4p$ (S $3p$)[71,72] and this interaction pushes the valence band upward, reducing effective $E_\mathrm{g}$,[73,74] as observed in CIGS.[66] In contrast, when the Cu content is low, the weaker $p$–$d$ interaction leads to the increase in $E_g$. In CZTSe and CZTS, therefore, $E_g$ is expected to change with the Cu composition [$x$ = Cu/(Zn + Sn)]. For the fabrication of CZTSe and CZTS solar cells, a variety of the $x$ values in a range of 0.77–0.86 (CZTSe)[23,24,26-30] and 0.76–0.98 (CZTS)[32,33,35-39] have been employed. However, the systematic variations of the CZTSe and CZTS dielectric functions with $x$ have not been determined. Thus, we modeled the increase in $E_g$ at low $x$ by shifting the CZTSe and CZTS dielectric functions toward higher energy using $\Delta E_g$ as an energy-shift parameter. In this case, however, based on the well-known sum rule,[75] the relation of $\int E\varepsilon_2(E)dE = \text{const.}$ needs to be satisfied. Thus, if the $\varepsilon_2$ spectrum is shifted toward higher energy by $\Delta E_g$, the $\varepsilon_2$ amplitude reduces by a factor of $f = E/(E + \Delta E_g)$ because $\int E\varepsilon_2(E)dE = \int (E+\Delta E_g)\varepsilon_2(E+\Delta E_g)dE$. Under these assumptions, the $\varepsilon_2$ spectrum of CZTSe layers can be expressed as

$$\varepsilon_2(E) = f\varepsilon_{2,\text{CZTSe}}(E - \Delta E_g) \tag{10}$$

In this equation, $\varepsilon_{2,\text{CZTSe}}$ shows the $\varepsilon_2$ spectrum of the CZTSe dielectric function in Fig. 4(a) ($\varepsilon_{\text{CZTSe}} = \varepsilon_{1,\text{CZTSe}} - i\varepsilon_{2,\text{CZTSe}}$) and the term of $-\Delta E_g$ shows the spectrum shift toward higher energy. In Eq. (10), we obtain $\varepsilon_2(E) = \varepsilon_{2,\text{CZTSe}}(E)$ when $\Delta E_g = 0$ eV. From the $\varepsilon_2(E)$ deduced from Eq. (10), the $\varepsilon_1(E)$ is obtained further using the Kramers-Kronig relations.[59] In this $\varepsilon_1(E)$ calculation, we used an $\varepsilon_1$ offset value of 0.770, estimated from the dielectric function modeling assuming the Tauc-Lorentz transitions.[61] In Fig. 5, $\varepsilon$ values calculated from our model using $\Delta E_g$ = 0.15 and 0.30 eV are shown, together with $\varepsilon_{\text{CZTSe}}$ in Fig. 4(a) ($\Delta E_g$ = 0 eV). The effect of the $x$ in the CZTS can also be modeled by simply adopting the $\varepsilon_2$ spectrum of the CZTS ($\varepsilon_{2,\text{CZTS}}$) in Eq. (10), instead of $\varepsilon_{2,\text{CZTSe}}$.

On the other hand, when Se in CZTSe is replaced with S, $E_g$ widening occurs by the upward shift of the conduction band and downward shift of the valence band.[76] As mentioned earlier, the overall shape of $\varepsilon_{\text{CZTSe}}$ is quite similar to that of $\varepsilon_{\text{CZTS}}$. In fact, when $\varepsilon_{2,\text{CZTSe}}(E)$ is slided toward higher energy, all the transition energies agree well with those of $\varepsilon_{2,\text{CZTS}}(E)$. Accordingly, to express the dielectric function of CZTSSe alloys, we employed the energy shift model in which the dielectric function of an arbitrary composition is "synthesized" from two known dielectric functions with different compositions.[77] When the CZTS$_y$Se$_{1-y}$ dielectric function for an arbitrary S composition [$y = S/(S + Se)$] is calculated by this model, $\varepsilon_{CZTSe}(E)$ is slided toward higher energy, while $\varepsilon_{CZTS}(E)$ is moved toward lower energy, so that the $E_g$ position of these shifted spectra matches that of the target $y$ composition. From these shifted dielectric functions, the dielectric function of the CZTSSe ($\varepsilon_{CZTSSe}$) is approximated as a weighted average of $\varepsilon_{CZTSe}$ and $\varepsilon_{CZTS}$:

$$\varepsilon_{CZTSSe}(E) = (1 - y)\, \varepsilon_{CZTSe}(E_{shift,Se}) + y\, \varepsilon_{CZTS}(E_{shift,S}) \qquad (11)$$

where $\varepsilon_{CZTSe}(E_{shift,Se})$ and $\varepsilon_{CZTS}(E_{shift,S})$ denote the CZTSe and CZTS dielectric functions obtained after the energy-shift adjustments.

In the energy region near $E_g$, $\varepsilon_{2,CZTSe}(E)$ overlaps with $\varepsilon_{2,CZTS}(E)$ when $\varepsilon_{2,CZTSe}(E)$ is shifted by 0.34 eV toward higher energy. Thus, we express the $E_g$ position of the CZTS$_y$Se$_{1-y}$ as $E_g = 0.91 + 0.34y$ eV. In an ideal semiconductor alloy, $E_g$ changes linearly with the alloy composition, but non-linear $E_g$ variation (band gap bowing) also occurs in many semiconductor alloys. In CZTSSe, the effect of the band gap bowing has been reported to be small[76,78,79] and this effect is neglected in our model. As a result, if the $y$ value is selected, the $E_g$ position is determined, and $\varepsilon_{CZTSe}$ and $\varepsilon_{CZTS}$ are shifted toward the target $E_g$ value. From these shifted spectra, $\varepsilon_{CZTSSe}$ is then calculated according to Eq. (11). Figure 6 shows $\varepsilon_{CZTSSe}$ for different $y$ values deduced from the energy shift model. The Cu contents of $\varepsilon_{CZTSe}$ and $\varepsilon_{CZTS}$ in Fig. 4 are $x = 0.95$ (CZTSe)[61] and $x = 0.94$ (CZTS)[68] and thus the above model corresponds to $\varepsilon_{CZTSSe}$ of $x \sim 0.95$. In CZTSSe, the reduction of $x$ is also expected to shift the dielectric function toward higher energy, but this effect will be interpreted as an increase in the $y$ composition.

As known well, the free carrier absorption in TCO materials, such as ZnO:Al and In$_2$O$_3$:Sn (ITO), alters EQE spectra significantly and, for accurate EQE simulations, the free carrier absorption in the TCO layers needs to be expressed properly.[19] Quite fortunately, the free carrier absorption in TCO layers can be described perfectly from the Drude model

$$\varepsilon_{Drude}(E) = -A_D/(E^2 - i\Gamma_D E), \qquad (12)$$

where $A_D$ and $\Gamma_D$ are the amplitude and broadening parameters, respectively.[59] From these parameters, the optical carrier concentration ($N_{opt}$) and optical mobility ($\mu_{opt}$) can be estimated from the relations of $A_D = \hbar^2 e^2 N_{opt}/(m^* \varepsilon_0)$ and $\Gamma_D = \hbar e/(m^* \mu_{opt})$.[80]

Here, $m^*$ is the effective mass. It should be noted that $N_{opt}$ shows quite good agreement with the carrier concentration determined by the Hall measurement.[80] In contrast, $\mu_{opt}$ generally shows a much higher value than the mobility obtained from the Hall measurement, as $\mu_{opt}$ does not include the effect of grain boundary scattering.[80,81] In the actual EQE analyses for reported spectra, the parameter values of $A_D$ and $\Gamma_D$ were modified slightly in some cases to improve the EQE fitting. Moreover, in TCO layers, the $E_g$ position shifts toward higher energy by the increase in $N_{opt}$ (Burstein-Moss shift).[80,82-84] Accordingly, in some analyses, the energy position of the TCO interband transition was also adjusted slightly by shifting the TCO dielectric function.

For the EQE analyses of CZTSe, CZTS, and CZTSSe solar cells, we employed the reported dielectric functions of non-doped ZnO,[19] CdS,[19,62] Mo,[19] and MgF$_2$,[19] while the free carrier absorption and the Burstein-Moss shift in ZnO:Al (Ref. 19), ZnO:Ga (Ref. 80) and ITO (Ref. 80) layers were modified slightly in some cases. For the MoSe$_2$ layer in CZTSe and CZTSSe solar cells, the dielectric function reported in Ref. 63 was used, whereas we employed a reported MoS$_2$ dielectric function[85] for CZTS solar cells.

We performed the EQE analysis of a CdTe solar cell[48] using the dielectric functions of CdTe (Fig. 4) and ITO (Ref. 80). For the MgF$_2$ and CdS layers incorporated into the CdTe solar cell, the dielectric functions mentioned above were employed. In addition, for the carbon rear electrode of the CdTe solar cell, we used the dielectric function of graphite.[86] In the EQE analyses of CH$_3$NH$_3$PbI$_3$ hybrid perovskite solar cells,[49,50] on the other hand, the dielectric functions reported for CH$_3$NH$_3$PbI$_3$ (Fig. 4), SnO$_2$:F (Ref. 87), TiO$_2$ (Ref. 65) and Au (Refs. 86, 88) were employed.

**IV. RESULTS**

**A. Analysis of CZTSe solar cells**

Figure 7 shows the result of the EQE analysis performed for the CZTSe solar cell in Fig. 1 using the e-ARC method. For this calculation, the optical model of Fig. 2(b) was used. In Fig. 7, the experimental EQE spectrum is shown by the open circles and the calculated EQE spectrum obtained from the fitting analysis is indicated by the red line. The black lines represent the $R_{tex}$ and $A$ spectra obtained from the ARC method. In this analysis, the variation of $x$ in the CZTSe layer is taken into account by employing $\Delta E_g$. In Fig. 7, therefore, we carried out the EQE analysis using $L_C$ in Eq. (5) and $\Delta E_g$ in Eq. (10) as free parameters.

In this CZTSe solar cell, $A$ in the short $\lambda$ region is limited essentially by the light absorption in the ZnO:Al and CdS layers. Specifically, the sharp reductions in the EQE response observed at $\lambda < 400$ nm and $\lambda < 540$ nm correspond to the onsets of strong light absorption in the ZnO:Al and CdS layers, respectively. The EQE reduction at $\lambda = 400-540$ nm, induced by the CdS absorption, is governed primarily by the CdS thickness,[19] and the CdS layer thickness estimated from the TEM image (70 nm) provides excellent fitting in Fig. 7. From the EQE spectra, the CdS layer thickness can also be deduced rather easily.

The light absorption observed in the ZnO:Al at $\lambda > 450$ nm shows the contribution of free carrier absorption, which shows a notable increase at longer wavelengths.[59,80] In CZTSe solar cells, the EQE at $\lambda = 600–1000$ nm is limited mainly by $R_{tex}$ and the free carrier absorption. In Fig. 7, the Drude parameters of the ZnO:Al are $A_D = 0.869$ eV and $\Gamma_D = 0.111$ eV, which correspond to $N_{opt} = 1.8 \times 10^{20}$ cm$^{-3}$ and $\mu_{opt} = 35.0$ cm$^2$/(Vs), respectively. It has been reported that $A_D$ of TCO layers changes depending on the underlying structure[19] and total TCO thickness.[81]

In Fig. 7, the $A$ spectrum of the CZTSe layer, shown as the yellow-colored region, corresponds to the EQE spectrum calculated within the ARC method ($L_C = \infty$), but the experimental EQE spectrum shows quite low values particularly in the long $\lambda$ region. This result indicates the significant carrier loss within the CZTSe layer. In the analysis based on the e-ARC method, however, the calculated EQE spectrum provides excellent fitting to the experimental EQE spectrum when the parameter values of $L_C = 0.57$ $\mu$m and $\Delta E_g = 0.16$ eV are assumed. The $J_{sc}$ value calculated from this analysis is 31.6 mA/cm$^2$ and agrees perfectly with the experimental value. The relatively large $\Delta E_g$ of 0.16 eV is induced by a low $x$ of 0.82 used in the CZTSe absorber ($x = 0.95$ in Fig. 4). From the result of Fig. 7, the optical and recombination losses can be determined quantitatively.

In Fig. 7, the $L_C$ value in the CZTSe (~0.6 $\mu$m) is much smaller than the total thickness of the CZTSe absorber layer (~1.3 $\mu$m), and the EQE in the longer $\lambda$ region reduces by carrier recombination in the CZTSe bottom region. The limited carrier collection in CZTSe solar cells due to small $L_D$ has been pointed out previously.[15,16,23] In earlier studies on CZTSe solar cells, on the other hand, $E_g$ of the CZTSe layers is often estimated from the EQE response in the longer $\lambda$ region.[24-27] Nevertheless, such analyses may lead to the overestimation of $E_g$ if the EQE response in the longer $\lambda$ regime is limited by the carrier collection, as confirmed from Fig. 7.

We also performed the EQE analysis by applying $H_2(\lambda)$ of Eq. (7). Figure 8 shows the results of the EQE fitting analyses for the CZTSe solar cell obtained using different $H(\lambda)$ functions and optical models. In this figure, the red and blue lines represent the EQE results for the analyses using $H_1(\lambda)$ and $H_2(\lambda)$, respectively. The experimental data (gray open circles) and the calculation result of $L_C = 0.57$ $\mu$m (red line) correspond to those in Fig. 7. As mentioned earlier, $H_1(\lambda)$ and $H_2(\lambda)$ are equivalent when $L_D = 0$. Thus, the EQE result obtained from $W = 0.57$ $\mu$m and $L_D = 0$ $\mu$m using $H_2(\lambda)$ is identical to that calculated from $L_C = 0.57$ $\mu$m using $H_1(\lambda)$. In the EQE analyses by $H_2(\lambda)$, however, the two fitting parameters (i.e., $W$ and $L_D$) show a strong correlation and quite similar spectra are obtained in a range of $W$ = 0.30–0.57 $\mu$m with $L_D$ = 0 $\mu$m ($W$ = 0.57 $\mu$m) and $L_D$ = 0.50 $\mu$m ($W$ = 0.30 $\mu$m). In particular, we found a relationship of $L_C \sim W + L_D/2$ between the parameters of $L_C$ in $H_1(\lambda)$ and ($W$, $L_D$) in $H_2(\lambda)$. The reasonable fitting can also be obtained when the parameters of $W$ = 0 $\mu$m and $L_D$ = 1.5 $\mu$m are used, as shown in Fig. 8. Accordingly, it is rather difficult to determine $W$ and $L_D$ independently from the model of $H_2(\lambda)$, although the EQE fitting may improve slightly when these two parameters are employed in the analyses. The large correlation between $W$ and $L_D$ in the EQE analysis using $H_2(\lambda)$ has also been confirmed in our EQE analyses for CdTe and $CH_3NH_3PbI_3$ solar cells.

We also calculated the EQE spectrum by assuming that the total CZTSe thickness in the optical model is $L_C$ ($d_{CZTSe}$ = 0.57 $\mu$m) and this result is shown by the black line in Fig. 8. The overall trend of this EQE spectrum is similar to that obtained from $L_C$ = 0.57 $\mu$m. In other words, $L_C$ extracted from the EQE fitting analysis is roughly equivalent to the thickness of the absorber layer where the carrier extraction occurs predominantly. However, the absolute EQE values obtained from $d_{CZTSe}$ = 0.57 $\mu$m differ from those of $L_C$ = 0.57 $\mu$m in the long $\lambda$ region due to the effects of the carrier recombination and the stronger back-side reflection in thinner layers.

To visualize the carrier generation and collection within the CZTSe solar cell, we further calculated partial $A$ and EQE by dividing the CZTSe layer into 1-nm-thick sublayers, as implemented previously.[19] In Fig. 9(a), partial $A$ calculated for different depths from the CdS/CZTSe interface ($d$) and wavelengths is shown. The $A$ values are normalized by the maximum value and the region of $d$ > 0.95 $\mu$m corresponds to the CZTSe/void region in Fig. 2(b). If the partial $A$ spectrum obtained at different $d$ is integrated, the $A$ spectrum in Fig. 7 (yellow-colored region) can be obtained. The partial $A$ value is quite high at $d \sim 0$ μm and exhibits a rapid decay versus $d$ due to the strong light absorption in the CZTSe layer. In the region of $\lambda$ > 900 nm, however, the light absorption and the resulting carrier generation occur rather uniformly throughout the entire CZTSe layer due to the lower $\alpha$ values in this region. The oscillation pattern observed in this region represents the weak interference effect induced by the thin film structure.

In Fig. 9(a), the $H_1(\lambda)$ values calculated from $L_C$ = 0.57 $\mu$m are also shown. By multiplying partial $A$ by $H_1(\lambda)$, partial EQE within the CZTSe layer can be estimated [Fig. 9(b)]. If the partial EQE spectra in Fig. 9(b) are integrated, the EQE spectrum shown in Fig. 7 is obtained. The trends of partial $A$ and partial EQE are essentially the same at $\lambda \leq 600$ nm, since $H_1(\lambda) \sim 1$ in this region. In the longer $\lambda$ region, however, the carrier collection is hindered strongly by the recombination in the bottom region and the partial EQE reduces significantly. It should be emphasized that, at $\lambda \geq 900$ nm in Fig. 9(a), the holes and electrons generated near the CdS/CZTSe and CZTSe/MoSe$_2$ interfaces, respectively, need to travel the whole layer to generate the current in the device. Accordingly, short $L_C$ lowers partial EQE particularly in the longer $\lambda$ region.

In Fig. 9(c), partial EQE in the CZTSe layer is integrated toward the depth to deduce the contribution of partial EQE at each depth for $J_{sc}$. The $J_{sc}$ profile obtained from partial $A$ assuming $L_C = \infty$ is also shown in Fig. 9(c). It can be seen that the light absorption in the CZTSe occurs predominantly at $d < 200$ nm and $J_{sc}$ increases rapidly in this thickness region. In particular, CZTSe shows a quite high $\alpha$ value exceeding $1 \times 10^5$ cm$^{-1}$ at 2.0 eV ($\lambda = 620$ nm) and the resulting penetration depth ($d_p = 1/\alpha$) is only ~100 nm. Thus, the rapid rise of $J_{sc}$ up to 200 nm can be explained by the high $\alpha$ values in the CZTSe layer. In contrast, the increase in $J_{sc}$ is smaller in the CZTSe bottom region, as the carrier collection in this region is limited by the enhanced recombination in this solar cell. Moreover, as indicated by the dotted line in Fig. 9(c), $J_{sc}$ becomes almost constant in the region of $d \geq L_C$. The contribution of each $\lambda$ for $J_{sc}$ is also calculated for the cases of $L_C = \infty$ [Fig. 9(d)] and $L_C = 0.57$ $\mu$m [Fig. 9(e)]. When the $L_C$ value is limited, the $J_{sc}$ contribution from the longer $\lambda$ region is lost and $J_{sc}$ reduces by 6.5 mA/cm$^2$ in Fig. 9(e).

Figure 10 shows the normalized partial EQE of the CZTSe solar cell with $L_C = 0.57$ $\mu$m and this figure corresponds to partial EQE shown in Fig. 9(b). It can be confirmed from Fig. 10 that the light absorption at $\lambda < 700$ nm occurs predominantly near the CdS interface, whereas partial EQE is distributed uniformly toward the depth direction at $\lambda > 900$ nm. In the CZTSe solar cell, therefore, the carrier collection in this regime is quite important to gain sufficient $J_{sc}$.

By applying the e-ARC method, we have further analyzed the EQE spectra of various CZTSe solar cells fabricated by a coevaporation,[23,26,27] a solution-based processing using hydrazine,[24] and a selenization process.[25] Figure 11 summarizes the experimental EQE spectra of the reported CZTSe solar cells (open circles) and the results of our EQE analyses (solid lines). The EQE analyses performed for these solar cells are essentially similar to that in Fig. 7; the thicknesses of the solar-cell component layers were extracted from the descriptions and the SEM images in each reference and, in the EQE fitting, $L_C$ and $\Delta E_g$ of the CZTSe layers were varied. In the analyses, we also adjusted the TCO optical properties and the CdS thickness slightly. Although there is slight uncertainty for the TCO optical constants in the analyses of the reported EQE spectra, we confirmed that the TCO optical properties can still be deduced from the EQE response in the visible region ($\lambda \sim 600$ nm) and the accurate analysis of $L_C$ is possible.

For the EQE analysis of Ref. 23, we employed the experimental $R$ spectrum, instead of $R_{tex}$, to simplify the analysis. For the other EQE calculations, $R_{tex}$ calculated from the ARC method was employed. As confirmed from Fig. 11, the e-ARC method provides excellent fitting to all the EQE spectra, independent of the detailed solar cell structure and processing method.

The EQE spectrum of Ref. 23 is obtained from a record-efficiency CZTSe solar cell with a conversion efficiency of 11.6% ($J_{sc}$= 40.6 mA/cm$^2$, $V_{oc}$ = 423 mV, $FF$ = 0.673). This solar cell has a very uniform thin film structure with large CZTSe grains and shows the highest $L_C$ value of 2.3 $\mu$m. As shown in Fig. 11, with decreasing the $L_C$ value, the EQE response in the longer $\lambda$ region reduces gradually and $J_{sc}$ reduces from 40.6 mA/cm$^2$ (Ref. 23) to 27.0 mA/cm$^2$ (Ref. 27).

Other EQE spectra reported for CZTSe solar cells[16,28-30] have also been analyzed by the e-ARC method and excellent fittings, similar to those in Fig. 11, have been obtained for these CZTSe solar cells. Figure 12 summarizes (a) $J_{sc}$, (b) $V_{oc}$ and (c) efficiency of the CZTSe solar cells as a function of $L_C$ estimated from our EQE analyses. From the $\Delta E_g$ obtained from each analysis, we calculated $E_g$ of the CZTSe layer as $E_g$ = 0.91 + $\Delta E_g$ eV. In Fig. 12, we categorized the analysis results into two groups depending on the $E_g$ values of the CZTSe layers. In particular, the closed circles represent the analysis results for the CZTSe layers with relatively high $E_g$ values in a range of 1.03–1.08 eV, whereas the closed squares show those with the lower $E_g$ values ($E_g$ = 0.97–1.02 eV). The solid lines in Fig. 12(a) indicate the variations of $J_{sc}$ with $L_C$, calculated assuming $E_g$ = 0.97 eV (blue line) and 1.04 eV (red line) for the CZTSe absorber. For these simulations, we assumed the structure of Fig. 2(b) but with a uniform CZTSe layer (2 $\mu$m) without the CZTSe/void layer.

As confirmed from Fig. 12(a), simulated $J_{sc}$ reproduces the experimental result quite well. The experimental $J_{sc}$ value shows a higher value than simulated $J_{sc}$ when (1) $E_g$ of the CZTSe is lower, (2) the CZTSe layer is thicker, (3) the free carrier absorption in the TCO layer is smaller, and (4) the CdS layer is thinner, compared with our model structure. Quite interestingly, $V_{oc}$ of the CZTSe solar cells is rather independent of $L_C$ and, for the solar cells of $E_g$ = 1.03–1.08 eV, $V_{oc}$ shows an upper limit of ~425 mV at $L_C$ = 0.6–2.3 $\mu$m, as indicated by the dotted line in Fig. 12(b). When $E_g$ is lower, on the other hand, $V_{oc}$ reduces rather largely. Thus, CZTSe solar cells show higher performance when CZTSe layers with high $E_g$ are formed by reducing the Cu content, as pointed out previously.[89] The variation of the conversion efficiency with $L_C$ [Fig. 12(c)] is essentially governed by the change in $J_{sc}$, as $V_{oc}$ is constant when $E_g$ is high. For the further increase in the efficiency, the improvement of $V_{oc}$ is crucial.

**B. Analysis of CZTS and CZTSSe solar cells**

We have further analyzed the EQE spectra of CZTS and CZTSSe solar cells by applying the e-ARC method. Figure 13 shows the EQE analysis results for (a) CZTS and (b) CZTSSe solar cells. In this figure, the open circles show the experimental spectra reported for CZTS (Refs. 31-35) and CZTSSe (Refs. 17, 40-43) solar cells, whereas the solid lines represent the calculated EQE spectra. The EQE analysis procedure in Fig. 13 is identical to that of CZTSe in Fig. 11, except that the dielectric functions of the CZTS and CZTSSe are applied in the analyses. As a result, the EQE fitting analyses for the CZTS solar cells were carried out using the two main parameters of $L_C$ and $\Delta E_g$.

The CZTS solar cells in Fig. 13(a) were fabricated by coevaporation[31] and sulfurization processes.[32-35] For these solar cells, the $x$ values of 0.76–0.9 and the Zn/Sn ratios of 0.91–1.58 are applied,[32-35] which result in the $E_g$ values of 1.44–1.50 eV in our analyses. It can be seen from Fig. 13(a) that the e-ARC method provides excellent fitting to the EQE spectra of the CZTS solar cells and the EQE response in the longer $\lambda$ region improves systematically with increasing $L_C$. The EQE spectrum of Ref. 31 is obtained from a high efficiency CZTS solar cell with a conversion efficiency of 8.4% ($J_{sc}$ = 19.5 mA/cm$^2$, $V_{oc}$ = 661 mV, FF = 0.658). The $L_C$ values extracted from the CZTS solar cells are smaller than those of the CZTSe in Figs. 11 and 12. Moreover, at $\lambda$ = 500−600 nm, the experimental EQE values for Refs. 33 and 35 are slightly smaller, compared with the calculated EQE values, indicating that the carrier recombination occurs slightly near the CdS/CZTS interface.

The CZTSSe solar cells in Fig. 13(b) have been fabricated by a solution-based processing using hydrazine[17,40] and selenization/sulfurization processes.[41-43] In the EQE analysis of these solar cells, $\varepsilon_{CZTSSe}$ was calculated from Eq. (11) and the EQE fitting was performed mainly by using the $y$ composition of CZTS$_y$Se$_{1-y}$ and $L_C$. The $E_g$ value of the CZTS$_y$Se$_{1-y}$ can then be calculated from the relation of $E_g$ = 0.91 + 0.34$y$ eV mentioned earlier. The excellent EQE fitting in the CZTSSe solar-cell analyses also confirms the validity of the e-ARC method and the variation of the EQE spectra can be reproduced almost perfectly by the change in $L_C$. In the CZTSSe solar cells with low $L_C$ values,[42,43] however, the EQE fitting deteriorates at $\lambda$ = 500−700 nm, indicating the presence of the recombination in the front interface region.

For the CZTSSe solar cells, the low $y$ values of 0.25–0.40 and the Cu composition of $x$ = 0.8–1.0 were employed with Zn/Sn ratios of 1.0–1.2.[17,41-43] From the EQE analyses of these solar cells, we determined $E_g$ of the CZTSSe layers to be 1.08–1.10 eV. The larger $E_g$ values in the CZTSSe, compared with the CZTSe ($E_g$ = 0.91 eV), can be interpreted by the high $y$ and the low $x$ values. In our dielectric function modeling of CZTSSe, however, the increase in $E_g$ due to a low $x$ is neglected and the $E_g$ shift induced by the low $x$ is compensated by the increase in $y$. Thus, in the actual EQE analyses for the CZTSSe, we obtained $y$ values of 0.50–0.65, which are higher than the experimental $y$ composition (0.25–0.40).[17,42,43] The $y$ values extracted from our EQE analyses are compared further with the reported $(x, y)$ values, and we find that the $E_g$ shift caused by the change in $x$ ($\Delta x$) can be expressed as $\Delta E_g \sim 0.5\Delta x$ eV. Thus, the reduction in $x$ from 0.9 to 0.8 (i.e., $\Delta x = 0.1$) increases $E_g$ by ~50 meV.

We implemented similar EQE analyses for other CZTS solar cells[36-39] and CZTSSe solar cells.[44-47] Figure 14 summarizes (a) $J_{sc}$, (b) $V_{oc}$, (c) $FF$ and (d) efficiency for all the CZTSe (Refs. 16, 23-30), CZTS (Refs. 31-39) and CZTSSe (Refs. 17, 40-47) solar cells analyzed in this study as a function of $L_C$ extracted from the EQE analyses. The solid lines in this figure indicate the variations of $J_{sc}$ with $L_C$, calculated assuming the $E_g$ values of 1.04 eV (CZTSe), 1.08 eV (CZTSSe) and 1.46 eV (CZTS). For these simulations, we employed the structure of Fig. 2(b) with a uniform absorber layer (2 $\mu$m). Quite naturally, as $E_g$ of the absorber material decreases, the overall $J_{sc}$ value increases. The result for $J_{sc}$ [Fig. 14(a)] indicates that the $L_C$ values in the CZTS are smaller than those in the CZTSe and CZTSSe. Quite interestingly, for the variation of $V_{oc}$ with $L_C$, CZTS shows the larger increase, compared with the CZTSe and CZTSSe. In particular, $V_{oc}$ of the CZTSe is quite independent of $L_C$ in a wide range and increases only slightly by 83 mV when $L_C$ increases from 0.2 to 2.3 $\mu$m, whereas $V_{oc}$ in the CZTS improves more than 184 mV in a similar $L_C$ range. Accordingly, it appears that the limiting factors of $V_{oc}$ in these solar cells are different.

In the CZTSe and CZTSSe solar cells, $FF$ increases with increasing $L_C$. The low $FF$ value in the CZTS solar cells suggests the lower quality of the CZTS absorber layers. It can be seen from Fig. 14(d) that, among the CZTSe, CZTS and CZTSSe solar cells, the CZTSSe shows the highest conversion efficiency. Although the $L_C$ values of the CZTSe and CZTSSe layers are rather similar, the CZTSSe solar cell shows better performance due to higher $V_{oc}$. Furthermore, our analysis reveals that $V_{oc}$ of the CZTSe and CZTSSe solar cells shows a weak correlation with $E_g$; estimated $E_g$ of the best performing CZTSe solar cell in Fig. 14(b) ($V_{oc}$= 423 mV, $L_C$ =2.3 $\mu$m for Ref. 23) is 1.07 eV, which is comparable to $E_g$ = 1.13 eV obtained for the high efficiency CZTSSe solar cell ($V_{oc}$ = 513 mV, $L_C$ = 0.75 $\mu$m for Ref. 17). Thus, the incorporation of the S atoms appears to be effective in increasing $V_{oc}$, although analysis errors of ±0.05 eV are expected for the $E_g$ values deduced in our EQE analyses.

**C. Analysis of a CdTe solar cell**

We have analyzed the EQE spectrum of a reported CdTe solar cell that shows a conversion efficiency of 16.0% ($J_{sc}$=26.1 mA/cm$^2$, $V_{oc}$ = 840 mV, FF = 0.731).[48] The structure of this device is MgF$_2$/glass/ITO (200 nm)/CdS (50 nm)/CdTe (3.5 $\mu$m)/carbon electrode. The CdTe layer of this solar cell was processed by a standard close-spaced sublimation (CSS) method,[90,91] followed by the aqueous CdCl$_2$ and the subsequent thermal-annealing (420 $^o$C) treatments. In CdTe processes, the post CdCl$_2$/annealing treatment is quite important to increase CdTe grain size and thus to reduce the carrier recombination at grain boundaries.[91-94] So far, a high conversion efficiency exceeding 20% has been reported for a CdTe-based solar cell,[95] although the structure of this device is not clear. It should be noted that only limited EQE analyses[70,96-98] and optical simulations[99,100] have been made for CdTe photovoltaic devices.

Figure 15 shows the result of the EQE analysis for the reported CdTe solar cell.[48] The shape of the experimental EQE spectrum (open circles) is essentially similar to those of the CZTSe and CZTS, although the longer $\lambda$ response is limited in the CdTe solar cell due to larger $E_g$. In Fig. 15, the red line represents the EQE spectrum calculated from the e-ARC method, while the EQE spectrum of $L_C = \infty$ is indicated by the yellow-colored region. In this EQE analysis, the thicknesses of the ITO, CdS and CdTe layers were adopted from the description in Ref. 48, whereas the thickness of the anti-reflection MgF$_2$ layer was assumed to be 110 nm. In the EQE fitting analysis of Fig. 15, only $L_C$ and the free carrier absorption in the ITO layer (i.e., $A_D$) were adjusted. We implemented optimization of the $A_D$ value so that the EQE values at $\lambda$ = 600–700 nm match the experimental values and, as a result, we obtained $A_D$ = 1.20 eV with a fixed parameter of $\Gamma_D$ = 0.12 eV [$N_{opt}$ = 2.8 × 10$^{20}$ cm$^{-3}$ and $\mu_{opt}$ = 29.5 cm$^2$/(Vs)]. It can be seen that the calculated EQE spectrum agrees quite well with the experimental spectrum when $L_C$ = 1.1 $\mu$m. So far, for CdTe solar cells, the $L_D$ values in a range of 0.4–3.7 $\mu$m with $W$ = 0.2–5.5 $\mu$m have been reported.[96,97,101-103]

In Fig. 15, the experimental EQE is slightly smaller than the calculated EQE in a region of $\lambda$ = 520–570 nm, indicating the small carrier loss near the CdS/CdTe interface. It is now well established that the CdS$_x$Te$_{1-x}$ alloy is formed at the CdS/CdTe interface.[90,92] Thus, the slight EQE reduction suggests the carrier recombination within the CdSTe phase. On the other hand, the disagreement at $\lambda$ > 850 nm originates from the light absorption by the tail state. Although polycrystalline CdTe layers are expected to show strong tail absorption, we employed the CdTe dielectric function extracted from a single crystal[69] and the effect of the tail absorption was not considered in our analysis.

Quite surprisingly, $L_C$ of the CdTe solar cell is comparable to that of the CZTSe solar cells in Fig. 11 (~1 $\mu$m), even though the carrier recombination loss in the CdTe solar cell is rather small (1.0 mA/cm$^2$ in Fig. 15). This effect can be understood based on partial EQE of the CdTe solar cell. Figure 16 shows (a) partial $A$ for the ITO (200 nm)/CdS (50 nm)/CdTe (3.5 $\mu$m) and (b) partial EQE ($L_C$ = 1.1 $\mu$m) for the CdTe device in Fig. 15. When the partial $A$ values for the ITO, CdS and CdTe layers are integrated, those in Fig. 15 can be obtained, whereas partial EQE corresponds to the EQE spectrum in Fig. 15. In Fig. 16, the partial $A$ and EQE values are normalized by the maximum values obtained in each layer. As shown in Fig. 16(a), the strong light absorption occurs below $\lambda$ that corresponds to $E_g$ of each layer. Nevertheless, in the CdTe layer, most of the light is absorbed in the region quite close to the CdS/CdTe interface due to the high $\alpha$ values above $E_g$. Accordingly, even when the $L_C$ value is rather small ($L_C$ ~ 1 $\mu$m), the major carrier loss occurs only in a narrow $\lambda$ region of 800–850 nm. As a result, the influence of the carrier recombination is suppressed quite well in the case of the CdTe solar cell. This result represents the quite important phenomenon in photovoltaic devices; namely, carrier recombination dynamics are affected significantly by the $\alpha$ spectrum of a light absorber layer and the recombination loss within the bulk component reduces if $\alpha(\lambda)$ shows high values with sharp absorption edge. In this sense, $\alpha(\lambda)$ of CdTe has a more ideal shape, if compared with CZTSe and CZTS.

**D. Analysis of CH$_3$NH$_3$PbI$_3$ solar cells**

To study the working principles of hybrid perovskite solar cells, the optical and recombination losses in reported CH$_3$NH$_3$PbI$_3$ solar cells[49,50] were determined using the e-ARC method. In particular, we have analyzed the EQE spectra of the CH$_3$NH$_3$PbI$_3$ solar cells with and without an hole transport layer (HTL) to elucidate the effect of the HTL on the carrier recombination.

Figure 17 shows the optical models for CH$_3$NH$_3$PbI$_3$ solar cells (a) with an HTL (Ref. 49) and (b) with no HTL (Ref. 50), and the corresponding EQE analysis results are shown in (c) and (d), respectively. The solar cell in Fig. 17(a) has a structure of glass/SnO$_2$:F/TiO$_2$/CH$_3$NH$_3$PbI$_3$/polytriarylamine(PTAA)/Au. In this device, the PTAA layer corresponds to the HTL. In more conventional CH$_3$NH$_3$PbI$_3$ solar cells, mesoporous TiO$_2$ layers are inserted between the uniform (compact) TiO$_2$ and CH$_3$NH$_3$PbI$_3$ layers.[104-107] However, for the CH$_3$NH$_3$PbI$_3$ solar cell of Fig. 17(a), a new spin-coating process allows the removal of the mesoporous TiO$_2$ layer and a high efficiency of 17.6 % has been obtained ($V_{oc}$ = 1.1 V, $J_{sc}$ = 20.5 mA/cm$^2$ and FF = 0.78).[49] For this solar cell, the optical modeling becomes quite simple because of a flat device structure.

In the optical model of Fig. 17(a), a complicated structure of SnO$_2$:F/SiO$_2$/SnO$_2$/glass (TEC-8, Pilkington) has been simplified to a SnO$_2$:F (600 nm)/glass structure. The validity of this approximation has been confirmed previously.[65] From the SEM image reported in Ref. 49, the thicknesses of the compact TiO$_2$ and CH$_3$NH$_3$PbI$_3$ layers were determined to be 30 nm and 300 nm, respectively. Unfortunately, for the PTAA layer, the optical constants have not been reported. However, the PTAA shows light absorption only in a high energy region of $E \geq 2.9$ eV,[108] which is notably larger than $E_g = 1.61$ eV of CH$_3$NH$_3$PbI$_3$. In the device, therefore, the light absorption within the PTAA is essentially negligible. In fact, spiro-OMeTAD [2,2',7,7'-tetrakis-(*N*,*N*-di-*p*-methoxyphenylamine)9,9'-spirobifluorene], which is used widely as an HTL of hybrid perovskite solar cells,[105-107] shows the same absorption onset at $E = 2.9$ eV,[65,109] and the optical loss in this layer is confirmed to be quite small.[65] Accordingly, in our optical model, the thickness of the PTAA layer is assumed to be zero. It should be emphasized that, when the EQE is calculated by incorporating a spiro-OMeTAD layer as an HTL, we observed no major changes. Thus, the influence of the HTL on the EQE spectrum is quite minor.

For the EQE analysis of the CH$_3$NH$_3$PbI$_3$ solar cells, we further assumed the carrier recombination in the TiO$_2$/CH$_3$NH$_3$PbI$_3$ interface region. This recombination is modeled by simply dividing a CH$_3$NH$_3$PbI$_3$ bulk layer into two layers and treating a thin layer located at the TiO$_2$ interface as a "dead layer" that allows no carrier extraction [see Fig. 17(a)].

In Fig. 17(c), the $A$ spectra of each solar-cell component layer calculated using the optical model of Fig. 17(a) are shown. In this figure, $R_{tex}$ deduced using the ARC method is also indicated. The yellow-colored region in Fig. 17(c) represents the $A$ spectrum of the CH$_3$NH$_3$PbI$_3$, and the hatched-line region corresponds to the $A$ component of the dead (recombination) layer near the TiO$_2$. We found that the calculated EQE (red line) shows good agreement with the experimental result (open circles) when the thickness of the dead layer is 4 nm with $L_C = \infty$. This result indicates clearly that the dominant recombination within the CH$_3$NH$_3$PbI$_3$ solar cell occurs near the TiO$_2$/CH$_3$NH$_3$PbI$_3$ interface region and the carrier loss near the HTL interface is negligible. Although the thickness of the front dead layer is thin (4 nm), the recombination loss by this layer leads to the reduction of $J_{sc}$ by 0.9 mA/cm$^2$. The carrier recombination near the TiO$_2$ interface region has also been suggested in our previous study.[65]

In Fig. 17(c), on the other hand, the agreement between the experimental and calculated EQE spectra is rather poor at $\lambda$ = 720–800 nm. The deviation of the calculated spectrum in this region is caused primarily by calculation errors of $R_{\text{tex}}$. Specifically, when the layer thickness of solar cells is thin, the period of the optical interference versus $\lambda$ becomes larger. In this case, when the minimum positions in the calculated $R_{\text{flat}}$ spectrum are simply connected, experimental $R$ cannot be reproduced accurately in the $E_g$ region where $R$ shows the sharp increase due to the negligible light absorption at $E < E_g$. In fact, at $\lambda = 750$ nm in Fig. 17(c), $R_{\text{tex}}$ almost overlaps with experimental EQE (i.e., $R_{\text{tex}} + EQE \sim 100\%$), which is not possible due to the light absorption within the $SnO_2$:F and Au layers. The above problem can be avoided if the experimental $R$ spectrum obtained from an actual solar cell is used in Eq. (4). In Fig. 17(c), however, the error in the $R_{\text{tex}}$ calculation affects $J_{\text{sc}}$ only slightly (0.4 mA/cm$^2$) and the quantitative analysis can still be made.

The solar cell with no HTL shown in Fig. 17(b) has a structure of glass/$SnO_2$:F/compact $TiO_2$/mesoporous $TiO_2$-$CH_3NH_3PbI_3$/$CH_3NH_3PbI_3$/Au.[50] As reported previously,[65] the optical response within the mesoporous $TiO_2$-$CH_3NH_3PbI_3$ mixed-phase layer can be expressed by the two separate flat layers of the $TiO_2$ and $CH_3NH_3PbI_3$. In the optical model of Fig. 17(b), the $TiO_2$ volume fraction within the mesoporous layer is assumed to be 30% (porosity 70%), which is consistent with the porosity of 58-64% reported in earlier studies.[110,111] With this assumption, the thicknesses of the component layers were determined based on the SEM image of the solar cell and the description in Ref. 50. However, minor adjustments were made for the layer thicknesses in the actual analysis, and the layer thicknesses used in our EQE analysis are summarized in Fig. 17(b).

Figure 17(d) shows the result of the EQE analysis performed for the $CH_3NH_3PbI_3$ solar cell with no HTL. The experimental EQE spectrum of this solar cell (open circles) is quite different from that of the $CH_3NH_3PbI_3$ solar cell with the HTL; the overall EQE values are smaller and the EQE response in the longer $\lambda$ region reduces significantly. The EQE spectra similar to that in Fig. 17(d) have been reported for $CH_3NH_3PbI_3$ solar cells that have no HTL layers.[112-115]

The yellow-colored region in Fig. 17(d), which corresponds to $A$ of the $CH_3NH_3PbI_3$ layer, is essentially the same with that in Fig. 17(c), as the device structures are quite similar. However, the $A$ spectra of the $TiO_2$ and $SnO_2$:F layers change slightly as the $TiO_2$ thickness is thicker and the $SnO_2$:F thickness is thinner in Fig. 17(d). The red line in Fig. 17(d) represents the EQE spectrum calculated assuming the carrier recombination in the front and rear interface regions. From the EQE values in the short wavelength region, the thickness of the front recombination layer is determined to be 10 nm and the EQE reduction caused by this layer is indicated by the hatched-line region.

For EQE in the longer $\lambda$ region, we obtained the good fitting when $L_C = 280$ nm. The result of Fig. 17 indicates clearly that the presence of an HTL modifies the carrier recombination within the $CH_3NH_3PbI_3$ significantly.

Figure 18 shows partial EQE obtained at different depths from the $TiO_2/CH_3NH_3PbI_3$ interface in the $CH_3NH_3PbI_3$ solar cells (a) with the HTL and (b) with no HTL. These partial EQE values correspond to the EQE spectra shown as the red lines in Figs. 17(c) and 17(d), and the partial EQE values are normalized by the maximum values in each solar cell. It can be seen that, at $\lambda < 500$ nm, the light absorption and the resulting carrier generation occur in the region close to the $TiO_2/CH_3NH_3PbI_3$ interface. Accordingly, the presence of the dead layer near the $TiO_2$ interface reduces the short $\lambda$ response greatly even if the thickness of the dead layer is thin, as evidenced from the hatched regions in Figs. 17(c) and 17(d).

In the longer $\lambda$ region of 600−800 nm, on the other hand, the carrier generation occurs rather uniformly toward the depth due to the relatively low $\alpha$ values in this region. In this case, the photocarriers generated at the front and rear interfaces need to travel the distance that corresponds to the whole $CH_3NH_3PbI_3$ thickness. The good agreement between the experimental and calculated EQE at $\lambda > 600$ nm, observed for the $CH_3NH_3PbI_3$ solar cell with the HTL [Fig. 17(c)], ensures the long carrier diffusion length, which is at least comparable to the $CH_3NH_3PbI_3$ layer thickness.[65] When the HTL is not present, however, the collection of carriers is hindered strongly and partial EQE in the longer $\lambda$ region reduces. The results of Fig. 18 confirm that the effect of the carrier recombination within the $CH_3NH_3PbI_3$ layer appears primarily as the EQE response at $\lambda > 600$ nm.

## V. DISCUSSION

### A. Optical and recombination losses

Figure 19 summarizes the experimental EQE spectra for various solar cells: CZTSe (Fig. 7), CZTSSe [Ref. 17 in Fig. 13(b)], CIGSe (Ref. 19), CZTS [Ref. 31 in Fig. 13(a)], CdTe (Ref. 48 in Fig. 15) and $CH_3NH_3PbI_3$ [Ref. 49 in Fig. 17(c)]. In Fig. 19, the reduction of EQE at $\lambda = 400-540$ nm, caused by the parasitic absorption in the CdS layers, is observed in all the solar cells, except for the $CH_3NH_3PbI_3$ solar cell. The EQE response in the longer $\lambda$ region is determined by $E_g$, but reduces when the recombination is present. In each EQE spectrum, the maximum EQE in the visible region is essentially governed by the light reflection and the free carrier absorption in the front TCO. By providing an anti-reflection layer and suppressing the free carrier absorption, high EQE values of ~95% can be achieved, as shown in Fig. 19.

Figure 20 illustrates the optical and recombination losses deduced from our EQE analyses performed for the spectra shown in Fig. 19: (a) CIGSe, (b) CZTSe, (c) CZTSSe, (d) CZTS, (e) CdTe, and (f) $CH_3NH_3PbI_3$. In this figure, the numerical values represent the corresponding current densities in units of mA/cm$^2$. The maximum $J_{sc}$ value attainable under AM1.5G condition is also calculated from the $\lambda$ position of the absorption edge in each absorber layer, and the optical gain indicated in Fig. 20 represents the ratio of output $J_{sc}$ divided by the maximum attainable $J_{sc}$ value. The $J_{sc}$ contributions obtained assuming two sublayers (top and bottom layers) with an equal thickness are also indicated. For the CIGSe, the result is adopted from our earlier EQE analysis.[19] This CIGSe solar cell is fabricated by the three-stage coevaporation process and has a standard structure of MgF$_2$ (130 nm)/ZnO:Al (360 nm)/non-doped ZnO (50 nm)/CdS (45 nm)/CIGSe (1.8 $\mu$m)/MoSe$_2$ (10 nm)/Mo/soda-lime glass with a conversion efficiency of 16.7% ($J_{sc}$= 34.2 mA/cm$^2$, $V_{oc}$ = 674 mV, FF = 0.725).

In Fig. 20, the CIGSe and CdTe solar cells show high optical gains of ~80%, whereas the optical gains of the CZTSe, CZTSSe, CZTS, and $CH_3NH_3PbI_3$ solar cells are in a range of 51−70%. The optical losses in these solar cells can mainly be categorized by (1) the reflection loss, (2) the free carrier absorption in the front TCO, (3) the parasitic absorption in the CdS dead layer, and (4) the strong light absorption by the Mo back contact. The reflection loss increases when $E_g$ of an absorber layer is low, as the EQE spectral range becomes wider and the integrated reflection loss increases in this case. More specifically, the reflection losses in the CZTSe and CZTSSe solar cells ($E_g \sim 1.0$ eV) are 2.5–4.5 mA/cm$^2$, which are apparently larger than those in the CZTS and CdTe solar cells. The $CH_3NH_3PbI_3$ solar cell also shows a large reflection loss (3.2 mA/cm$^2$) due to the lack of an anti-reflection coating and the thin absorber-layer thickness (~300 nm).

The parasitic absorption by front TCO layers is a common problem for all thin-film solar cells. The optical loss in a front TCO layer tends to increase in solar cells having low-$E_g$ absorbers due to the widening of the EQE spectral range. Moreover, the free carrier absorption in TCO layers increases drastically at longer wavelengths.[59,80] Accordingly, when low-$E_g$ absorbers are applied, the suppression of reflection loss and absorption loss in the front TCO becomes critical in achieving high conversion efficiencies. The Al-doped ZnO layers generally show a low $N_{opt}$ value of $\sim 2 \times 10^{20}$ cm$^{-3}$ [$\mu_{opt}$ = 30−50 cm$^2$/(Vs)][19,81] and the typical thickness of the ZnO:Al (~300 nm) results in the absorption loss of ~3 mA/cm$^2$ [see Figs. 20(a) and 20(b)]. In the CZTSSe solar cell in Fig. 20(c), the TCO absorption is suppressed by incorporating a thin ITO layer (50 nm).[116] In the CdTe solar cell, the optical loss caused by the front ITO is only 0.7 mA/cm$^2$ even though the ITO thickness is 200 nm. This small optical loss in the front TCO originates from the large $E_g$ value of CdTe, which limits the effect of the free carrier absorption. In contrast, the CH$_3$NH$_3$PbI$_3$ solar cell having a larger $E_g$ value shows the relatively large front TCO loss. This is caused by strong free carrier absorption originating from a thick TCO layer (600 nm) with a high $N_{opt}$ of $5 \times 10^{20}$ cm$^{-3}$ (TEC glass).[117,118] As reported previously,[119-121] the parasitic light absorption in the front TCO can be reduced by employing high-mobility TCO layers, which exhibit quite low free carrier absorption.

The optical loss induced by the CdS layer (~50 nm) is ~3 mA/cm$^2$, as observed for CIGSe, CZTSe, CZTS and CdTe solar cells in Fig. 20. In the CZTSSe solar cell [Fig. 20(c)], the optical loss in the CdS is reduced by incorporating a quite thin CdS layer (25 nm).[116] In high-efficiency CdTe solar cells, the parasitic absorption in the CdS layer is also reduced by adopting a thin layer.[94,122] In conventional CdTe solar cells, however, $V_{oc}$ shows a sharp reduction when the CdS thickness is reduced below 60 nm.[94,122] Quite fortunately, it has been found that such $V_{oc}$ reduction can be suppressed by incorporating a high-resistive TCO layer at the front TCO/CdS interface.[94,122] In a record efficiency CdTe-based solar cell (21.5%), the parasitic absorption in the CdS-absorption region is suppressed quite well.[95]

In the CIGSe and CZT(S)Se solar cells, a strong optical loss occurs by the parasitic absorption in the Mo layer. This effect can be attributed to the inherent problem of Mo: i.e., low reflectivity of Mo particularly in the longer $\lambda$ region.[19,123-125] The strong parasitic absorption in the Mo is evident in the EQE analysis result for the CZTSe solar cell (Fig. 7). The optical loss induced by the Mo increases when a thinner absorber is employed, as confirmed for the CZTS solar cell in Fig. 20(d). For CIGSe solar cells, the application of highly reflective metals, such as Au, Ag, Cu and Al, has been rather difficult in part due to a high processing temperature of the CIGSe absorber, although Au back electrodes have been employed for CIGS solar cells fabricated by a lift-off process.[125] For CIGSe and CZT(S)Se solar cells, the formation of more ideal back-reflector structures is preferable for the further increase in $J_{sc}$.

In all the solar cells, the $J_{sc}$ contribution of the top sublayer is significant, compared with that of the bottom sublayer. In the CdTe solar cell, the $J_{sc}$ contribution of the bottom sublayer is quite small due to the strong light absorption within the CdTe absorber layer, as confirmed from Fig. 16(a). Thus, the thickness of the CdTe layer could be reduced down to ~1 $\mu$m without deteriorating the efficiency to lower the production cost. However, the backside recombination may influence the performance in this case (see Sec. V B). In the case of the $CH_3NH_3PbI_3$ solar cell, however, $J_{sc}$ generated in the bottom region is still important due to the thin thickness of the $CH_3NH_3PbI_3$ absorber. It should be emphasized that the $CH_3NH_3PbI_3$ solar cell shows quite small overall optical losses. In particular, the parasitic absorption within the $TiO_2$ and HTL layers is negligible and the major optical loss occurs only in the front TCO layer, which can be optimized further. Thus, a quite high conversion efficiency reported for $CH_3NH_3PbI_3$ solar cells (~20%) can be understood partly by the low parasitic absorption within the solar-cell component layers.[65]

**B. Carrier recombination**

Although our model for the carrier recombination is extremely simple, almost all the EQE spectra reported for CZTSe, CZTS, CZTSSe, CdTe and $CH_3NH_3PbI_3$ solar cells can be analyzed using the identical formula of $H_1(\lambda)$. In some cases, however, the model of $H_1(\lambda)$ may not be appropriate, and such limitations need to be clarified further. Unfortunately, our EQE analyses performed for various solar cells do not allow the determination of the recombination mechanism, but our analyses provide the quantitative results for (1) the total $J_{sc}$ loss induced by the carrier recombination and (2) the thickness region where the carrier loss occurs predominantly (i.e., near front or rear interface region). As mentioned above, $L_C$ extracted from the e-ARC method is a reference value and does not correlate directly with $L_D$ and $W$. Accordingly, we emphasize that the discussion concerning recombination mechanisms described here is qualitative.

Our EQE analyses reveal that the recombination losses are well suppressed in CdTe and $CH_3NH_3PbI_3$ solar cells, while the CZTSe and CZTSSe solar cells show relatively large recombination losses. On the other hand, state-of-the-art CIGSe solar cells show almost 100% carrier collection with negligible carrier recombination.[19] The recombination losses in the various solar cells of Fig. 20 can be interpreted by the corresponding band diagrams. Figure 21 summarizes the schematic band diagrams of (a) CIGSe, (b) CZTSe, (c) CdTe and (d) $CH_3NH_3PbI_3$ solar cells. These diagrams are drawn by referring to those reported earlier for CIGSe solar cells,[126-130] CZTSe solar cells,[131-134] and CdTe solar cells,[98,135-137] whereas the band diagrams of $CH_3NH_3PbI_3$ solar cells in Fig. 21(d) represent those proposed in this study. For the CIGSe and CZTSe solar cells, however, the band diagram near the $MoSe_2$ is not shown, as the band diagram at the $MoSe_2$ rear interface remains controversial.[129,138] In Fig. 21, the band diagrams are shown by using a consistent energy scale, and the rough scales of $L_C$, $L_D$ and $W$ are also indicated. The carrier recombination processes that occur in the rear interface, bulk and front interface regions are also illustrated as *A*, *B*, and *C* in Fig. 21, respectively.

As known well, the CIGSe, CZTSe and CdTe exhibit *p*-type conductivity and the depletion region, formed with the *n*-type CdS, extends deeper into the *p*-type absorber layers.[126-137] In CIGSe and CZTSe solar cells, the conduction band offset ($\Delta E_C$) at the CdS/non-doped ZnO is 0.2 eV.[128,130,133,134] In the case of CIGSe solar cells, to suppress the carrier recombination at both front and rear interfaces, the V-shaped Ga profile is generally formed toward the growth direction by a three-stage process, in which Cu, In, Ga and Se elemental sources are supplied with different sequences.[139-141] In the resulting CIGSe layer, $E_g$ (Ga content) is low at the valley position, typically located at 0.5 $\mu$m from the CdS/CIGSe interface,[19,142] while $E_g$ increases toward the front and rear interfaces. In CIGSe alloys, the energy position of the conduction band minimum (CBM) shifts upward with increasing the Ga content.[143,144] Thus, the V-shaped Ga profile predominantly modifies the CBM position. In the band diagram of Fig. 21(a), the Ga-profile of the actual CIGSe solar cell [Fig. 20(a)] is reproduced and $\Delta E_C$ at the CdS/CIGSe interface becomes zero when a Ga composition at the front interface is 40 %.[145] Moreover, from the $E_g$ values at the valley and rear positions, the conduction band barrier is estimated to be 0.14 eV.[19] Although this barrier height is rather small, the pseudo-potential generated by the Ga-grading suppresses the rear-interface recombination quite effectively.[127,129,140,141,146] Moreover, the light absorption in the CIGSe layer (~2 $\mu$m) occurs primarily in the 1-$\mu$m-thick region from the CdS/CIGSe interface. From this result, we concluded previously that the CIGSe bottom layer (1 $\mu$m) with higher Ga contents plays a dominant role as a back-surface field (BSF) layer.[19]

In CZTSe solar cells [Fig. 21(b)], on the other hand, a quite large $\Delta E_C$ value at the CdS/CZTSe interface (0.48–0.6 eV) has been reported.[131,132] Moreover, in this solar cell, the flat-band formation is expected to occur as no intentional composition modulation is made within the CZTSe layer. Earlier studies on CZTSe and CIGSe solar cells show that the hole carrier concentration ($N_p$) of a CZTSe layer is $3 \times 10^{16}$ cm$^{-3}$ with $W = 0.2$ $\mu$m, which are comparable to the values of $N_p = 2 \times 10^{16}$ cm$^{-3}$ and $W = 0.4$ $\mu$m in a CIGSe.[147] In fact, in the device simulations for CIGSe[130,148] and CZTSe[133,134] solar cells, $N_p$ of ~$1 \times 10^{16}$ cm$^{-3}$ is generally assumed. Consequently, the $W$ values of the CIGSe and CZTSe solar cells are similar. Accordingly, the large difference in $L_C$ between the CIGSe ($L_C = \infty$) and CZTSe ($L_C = 0.57$ $\mu$m in Fig. 7) is likely caused by the variation in $L_D$, as illustrated in Figs. 21(a) and 21(b).

There are two possible recombination passes that are expected to lower $L_D$ in CZTSe solar cells: i.e., the recombination (1) at the CZTSe/MoSe$_2$ interface and (2) within the CZTSe bulk layer [i.e., *A* and *B* in Fig. 21(b)]. Unfortunately, from our EQE analyses, these recombination passes cannot be distinguished. However, since there is no BSF structure in CZTSe solar cells, the carrier recombination at the rear interface is possible. Moreover, in CZTS solar cells, the phase separation of CZTS into the $Cu_2SnS_3$ (CTS) and ZnS components occurs at the CZTS/$MoS_2$ rear interface.[149] This result implies that the formation of the CTSe and ZnSe secondary phase may also occur at the CZTSe/$MoSe_2$ interface. Since CTS and CTSe have lower $E_g$ values, compared with the CZTS and CZTSe, respectively, such compositional change likely leads to the severe recombination. The low $V_{oc}$ value in the CZTSe solar cells, which is independent of $L_C$ [see Fig. 14(b)], also suggests the interface-limited performance of CZTSe solar cells. In fact, when a TiN barrier layer is provided on the Mo substrate, the EQE response in the longer $\lambda$ region and the PL life time improve significantly.[54]

However, the possibility of the intense carrier recombination in the CZTSe grain boundary region cannot be ruled out. In particular, the high-resolution compositional analysis performed for CZT(S)Se layers revealed the higher Cu contents in the grain boundary region,[24,149] which likely enhance the CT(S)Se formation. Although scanning-probe microscopy (SPM) measurements confirm the enhanced carrier extraction from the grain boundary region in CZTSSe layers,[150] the formation of larger CZTSe grains has been quite effective in improving $L_C$.[23] Accordingly, it appears that the performance of the CZTSe solar cells is limited by intense carrier recombinations in both bulk and rear interface regions. Just recently, however, from bias-voltage-dependent EQE analyses and simulations, it has been reported that the carrier recombination in CZTSe[18] and CZTSSe[151] solar cells occurs predominantly at the band tail states and the conversion efficiency of the CZTSe-based solar cells is limited essentially by the formation of the tail states.[18,151]

In CdTe solar cells [Fig. 21(c)], on the other hand, the Fermi level ($E_F$) in the CdS layers locates near the CBM[98,135-137] and the CdS/CdTe interface shows the cliff-type discontinuity ($\Delta E_C$ = 0.2 eV)[152] due to larger $E_g$ of CdTe. For *p*-type CdTe layers, there have been no simple metal electrodes that provide ohmic contacts and the carbon electrode with excessive Cu is used as a rear electrode. The presence of the Cu atoms is important to lower the height of the Schottky barrier formed at the CdTe/C interface,[135] although the extensive diffusion of the Cu atoms into the CdTe absorber limits the stability of the devices.[135,153,154] It has been confirmed that the presence of the Schottky barrier at the CdTe rear interface reduces the current in the forward bias region.[136]

For CdTe solar cells, CdTe layers with a low $N_p$ value of $10^{14}$–$10^{15}$ cm$^{-3}$ have been applied.[98,136,137] This $N_p$ value is much lower than those of the CIGSe and CZTSe ($N_p$

~$10^{16}$ cm$^{-3}$). As a result, the depletion region in the CdTe extends far deeper into the CdTe layer ($W$ = 1–3 $\mu$m),[137,155] if compared with the CIGSe and CZTSe solar cells. In the CdTe solar cells, however, the light penetration at $\lambda$ < 800 nm occurs within 1 $\mu$m from the CdS/CdTe interface due to the high $\alpha$ values, as shown in Fig. 16(a). Accordingly, in the case of CdTe solar cells, most of the carriers are generated within the thick depletion region and the electric-field-assisted carrier collection occurs,[137,155] which is quite beneficial to increase the EQE response at longer $\lambda$. As a result, we observe $L_C \sim W$ in CdTe solar cells, whereas the relation of $L_C > W$ is confirmed for CIGSe and CZTSe solar cells. Thus, the contribution of $L_D$ is considered to be more important in the CIGSe and CZT(S)Se solar cells. In the CdTe solar cells, however, low $N_p$ leads to a small $E_F$ shift within the CdTe layer, which in turn reduces the built-in potential.[137,156] In fact, for the CdTe solar cell in Fig. 20(e), the $V_{oc}$ loss, defined by $V_{loss}$ = $E_g/e - V_{oc}$, is 650 mV and this $V_{loss}$ is larger than those observed in the CIGSe ($V_{loss}$ = 451 mV)[19] and CH$_3$NH$_3$PbI$_3$ (510 mV)[49] solar cells. On the other hand, the high-efficiency CZTSe,[23] CZTSSe,[17] and CZTS (Ref. 31) solar cells show $V_{loss}$ values of 647, 617 and 799 mV, respectively, if the $E_g$ values obtained from our EQE analyses are applied. To suppress the $V_{oc}$ loss in CdTe solar cells, the introduction of a ZnTe BSF layer at the CdTe/C interface has been proposed[137,157,158] and performed.[159,160] The recombination loss of 1.0 mA/cm$^2$ in the CdTe bottom region could also be eliminated by providing the ZnTe BSF layer at the rear interface.[157]

For CdCl$_2$-treated CdTe solar cells, the effective carrier collection from the grain boundary region has been confirmed from EBIC[2,161,162] and scanning capacitance microscopy.[163,164] Nevertheless, earlier studies on CdTe solar cells pointed out that $V_{oc}$ is still limited by the leakage current in the grain boundary region.[164,165] Indeed, in a single-crystal CdTe solar cell, a quite high $V_{oc}$ value of 929 mV ($V_{loss}$ = 561 mV) has been obtained at a similar $N_p$ value of ~6 × 10$^{15}$ cm$^{-3}$.[156] Thus, for the increase in $V_{oc}$ of CdTe solar cells, the formation of larger grain is important, in addition to high $N_p$. In the CIGSe, the growth of large grains (1–2 $\mu$m) has been realized by utilizing a liquid Cu-Se phase generated during the three-stage process,[140] whereas the grain size in the CZTSe has been improved by NaF post treatments.[23] Accordingly, in CIGSe, CZTSe and CdTe solar cells, the suppression of the grain boundary formation has been critical to obtain higher efficiencies.

In Fig. 21(d), the band diagrams for CH$_3$NH$_3$PbI$_3$ solar cells with an HTL (spiro-OMeTAD) and with no HTL are shown. The $\Delta E_C$ value at the TiO$_2$/CH$_3$NH$_3$PbI$_3$, determined from the direct and inverse photoemission spectroscopies, is ~0.1 eV,[166,167] and $E_F$ of the TiO$_2$ locates near the CBM[166] due to a high carrier concentration of 10$^{19}$

cm$^{-3}$ in this layer.[168] For the spiro-OMeTAD in Fig. 21(d), the energy positions of the highest occupied molecular orbital (HOMO) and lowest unoccupied molecular orbital (LUMO) are shown with an energy separation that corresponds to $E_g$ (2.9 eV).[65] When the spiro-OMeTAD is doped with Li, $E_F$ shifts toward the HOMO,[169] and the energy difference between the HOMO of the Li-doped spiro-OMeTAD and the valence band maximum of CH$_3$NH$_3$PbI$_3$ is reported to be ~0.2 eV from the ultraviolet photoelectron spectroscopy (UPS).[167] In Fig. 21(d), the energy offsets at the front and rear interfaces of the CH$_3$NH$_3$PbI$_3$ are reproduced by using the above experimental values. It can be seen that, in CH$_3$NH$_3$PbI$_3$ solar cells, the electrons and holes generated within the absorber layer are collected in the front and rear electrodes without any potential barriers.[105]

Unfortunately, for the electric-field distribution within the CH$_3$NH$_3$PbI$_3$ layer, quite different results have been reported[5,170-173] and the band diagrams proposed for the hybrid perovskite solar cells have been highly controversial.[4,170,171,174-177] It is generally assumed that the operation of CH$_3$NH$_3$PbI$_3$ solar cells occurs by the formation of a $p$–$i$–$n$ type structure.[5,173-178] Indeed, when EBIC measurements were implemented for CH$_3$NH$_3$PbI$_{3-x}$Cl$_x$ solar cells having the thick absorber layers (1.5 $\mu$m), the presence of two intense electric-field regions was confirmed near the TiO$_2$/CH$_3$NH$_3$PbI$_{3-x}$Cl$_x$ and CH$_3$NH$_3$PbI$_{3-x}$Cl$_x$/HTL interfaces.[4] Nevertheless, other studies based on Kelvin force microscopy reported that the electric field concentrates at the TiO$_2$/CH$_3$NH$_3$PbI$_3$ (or CH$_3$NH$_3$PbI$_{3-x}$Cl$_x$) interface[170,172] and several studies concluded that the hybrid perovskite solar cell operates as a simple $p$–$n$ junction device.[115,170-172]

This controversy can be interpreted by the carrier type of CH$_3$NH$_3$PbI$_3$ layers. To date, $n$-type conductivities of CH$_3$NH$_3$PbI$_3$ layers have been confirmed by Seebeck[179] and Hall[180] measurements. In solution-based processing of CH$_3$NH$_3$PbI$_3$, however, the carrier type changes from $p$-type to $n$-type with increasing PbI$_2$/CH$_3$NH$_3$I precursor ratio.[180] Thus, CH$_3$NH$_3$PbI$_3$ layers show the $n$-type carrier conduction when the PbI$_2$-rich phase is formed, although CH$_3$NH$_3$PbI$_3$ single crystals exhibit the $p$-type character with a low $N_p$ value of 9 × 10$^9$ cm$^{-3}$.[181] The earlier UPS measurement further confirmed that, when a CH$_3$NH$_3$PbI$_3$ layer is fabricated by a standard condition (PbI$_2$/CH$_3$NH$_3$I = 1), $E_F$ shifts toward CBM with an electron concentration of 3 × 10$^{17}$ cm$^{-3}$.[180]

Our EQE analyses in Fig. 17 strongly support the formation of the $n$-type CH$_3$NH$_3$PbI$_3$. In particular, when the CH$_3$NH$_3$PbI$_3$ layer shows $n$-type conductivity, the electric field is expected to concentrate at the CH$_3$NH$_3$PbI$_3$($n$)/HTL($p$) interface with a weak electric field at the TiO$_2$($n$)/CH$_3$NH$_3$PbI$_3$($n$) interface, as schematically shown in Fig. 21(d). Thus, the relatively strong front-carrier recombination near the TiO$_2$ interface [i.e., recombination pass C in Fig. 21(d)] could be explained by the weaker electric field in this region. In contrast, the recombination in the solar-cell bottom region is suppressed almost completely most likely due to the presence of the strong field near the HTL. Based on the above result, we propose that, in CH$_3$NH$_3$PbI$_3$ solar cells, the electric-field-assisted carrier collection ($L_C \sim W$) occurs by the formation of a $p$–$n^-$–$n$ type structure (or $n$–$n^-$–$p$ from the light illumination side).

This situation changes significantly when an HTL is not present. In this case, the flat-band formation is expected to occur by the elimination of the strong electric field in the bottom region. In this device, the photocarrier collection could be limited by the carrier diffusion, similar to the case of the CZTSe solar cell in Fig. 21(b), and the intensified recombination in the rear interface region lowers the EQE response in the longer $\lambda$ region, as pointed out previously.[114] In fact, the previous EBIC study confirmed that the strong electric field present at the CH$_3$NH$_3$PbI$_{3-x}$Cl$_x$/HTL interface disappears completely when the HTL is removed from the device.[4] Thus, lower $V_{oc}$ observed in CH$_3$NH$_3$PbI$_3$ solar cells with no HTL[112-114,182] can be interpreted by the smaller built-in potential. In contrast, the presence of the HTL does not affect the electric-field distribution within the CH$_3$NH$_3$PbI$_3$ when the $p$–$n$ junction is formed.[172] The above evidences lead us to conclude that the strong electric field is generated at the CH$_3$NH$_3$PbI$_3$/HTL for the $n$-type CH$_3$NH$_3$PbI$_3$, whereas the electric field concentrates at the CH$_3$NH$_3$PbI$_3$/TiO$_2$ for the $p$-type CH$_3$NH$_3$PbI$_3$. Nevertheless, the drastic reduction of the EQE response by the removal of an HTL has commonly been observed in CH$_3$NH$_3$PbI$_3$ solar cells.[112-115] Accordingly, we believe that CH$_3$NH$_3$PbI$_3$ layers incorporated into high-efficiency hybrid perovskite solar cells basically have $n$-type character.

Earlier DFT studies further revealed that mid-gap states are not created by the vacancy/interstitial-type defects in CH$_3$NH$_3$PbI$_3$,[183-185] although the formation of complex defects is still possible.[186] Moreover, in the grain boundary region of CH$_3$NH$_3$PbI$_3$, the PbI$_2$-rich phase having larger $E_g$ is formed, suppressing the carrier recombination in the grain boundary region effectively.[187] The efficient carrier extraction from CH$_3$NH$_3$PbI$_3$ grain boundary regions has also been reported by SPM[188] and EBIC.[189] In CH$_3$NH$_3$PbI$_3$ solar cells, therefore, the carrier recombination is expected to occur predominantly in the front and rear interface regions with limited recombination within the bulk region. However, it has recently been reported that the grain size of CH$_3$NH$_3$PbI$_3$ affects $V_{oc}$.[190] In this point of view, CH$_3$NH$_3$PbI$_3$ solar cells show a feature of polycrystalline solar cells and are similar to CIGSe, CZT(S)Se and CdTe solar cells.

The EQE analyses performed for CZT(S)Se, CdTe and $CH_3NH_3PbI_3$ solar cells in this study consistently suggest that the intense carrier recombination occurs when the flat band is formed. This is in sharp contrast to crystalline Si-based solar cells, in which a quite long carrier life time (~1 ms)[191] results in a quite long $L_D$ (~2000 $\mu$m) of photo-generated carriers. Accordingly, in thin-film solar cells with limited carrier life time and $L_D$, the electric-field-assisted carrier collection is of paramount importance to realize high conversion efficiencies.

## C. Effect of $L_C$ on $J_{sc}$

Figure 22 shows the variation of $J_{sc}$ with $L_C$ for various solar cells calculated from the e-ARC method. For all the solar cells, we assumed an identical solar cell structure consisting of the $MgF_2$ (130 nm)/ZnO:Al (360 nm)/ZnO (50 nm)/CdS (45 nm)/absorber (2.0 $\mu$m)/Mo, except for CdTe and $CH_3NH_3PbI_3$ solar cells. For the CdTe and $CH_3NH_3PbI_3$ solar cells, the structures of Fig. 15 (CdTe) and Fig. 17(a) ($CH_3NH_3PbI_3$) were adopted, but we performed the calculation using an identical absorber thickness of 2.0 $\mu$m to compare the intrinsic properties of the absorber layers. In the calculation of the $CH_3NH_3PbI_3$ solar cell, the thickness of the front dead layer is assumed to be zero. For the dielectric functions of the absorbers, the optical data of Fig. 4 were employed, although $E_g$ is increased slightly for the CZTSe ($E_g$ = 1.0 eV) and the CZTS ($E_g$ = 1.45 eV) based on the results of Fig. 14.

In Fig. 22, $J_{sc}$ is higher in the solar cell with a low-$E_g$ absorber and $J_{sc}$ increases rapidly with increasing $L_C$. The variations of $J_{sc}$ with $L_C$ are essentially similar in all the solar cells, and $J_{sc}$ tends to saturate at $L_C$ > 1.0 $\mu$m. Nevertheless, the $J_{sc}$ values of CdTe and $CH_3NH_3PbI_3$ solar cells show almost complete saturation at $L_C$ > 1.0 $\mu$m. This trend originates from the sharp absorption features of these absorbers, as confirmed in Fig. 4. Consequently, in CdTe and $CH_3NH_3PbI_3$ solar cells, $J_{sc}$ is not influenced by the recombination as long as the condition of $L_C$ > 1.0 $\mu$m is satisfied. Thus, the dispersion of the $\alpha$ spectrum is quite important in interpretating the carrier generation/recombination. For solar cells, it is preferable if a light absorber has high $\alpha$ values near $E_g$ with a low Urbach energy (sharp absorption onset). The CdTe and $CH_3NH_3PbI_3$ semiconductors appear to fulfill these conditions relatively well.

# VI. CONCLUSION

We developed a general EQE analysis method that allows the quantitative description of the optical and carrier recombination losses within thin-film solar cells. In the developed method, the light scattering caused by submicron textures is expressed by imposing anti-reflection condition in the calculation of the optical admittance method and the carrier recombination is further modeled based on exponential-decaying carrier extraction from the absorber interface using $L_C$ as a sole parameter. Although our EQE analysis method is quite simple, it is sufficient to determine optical and recombination losses quantitatively. In particular, if the optical constants and thickness of each layer are known, the EQE spectrum can be calculated quite easily without considering geometrical structures.

The developed method was applied to variety of solar cell devices fabricated by coevaporation, sputtering, closed-space sublimation and solution-based processes, and provides almost perfect fitting to numerous EQE spectra reported earlier. From our EQE analyses, the $L_C$ values in CZTSe, CZTS, CZTSSe, CdTe and $CH_3NH_3PbI_3$ solar cells were estimated. We found that $J_{sc}$ in the CZTSe and CZTSSe solar cells is limited severely by low $L_C$ values, originating from the intense recombination in the absorber bottom region. In contrast, CdTe and $CH_3NH_3PbI_3$ solar cells show small recombination losses. Finally, we emphasize that the developed method allows the fast and easy analysis of EQE spectra obtained from various thin-film solar cells without the requirement of intensive computer modeling for surface and interface structures. Accordingly, the EQE analysis technique developed in this study provides an ideal method for determining various limiting factors in thin-film photovoltaic devices.

**Figures**

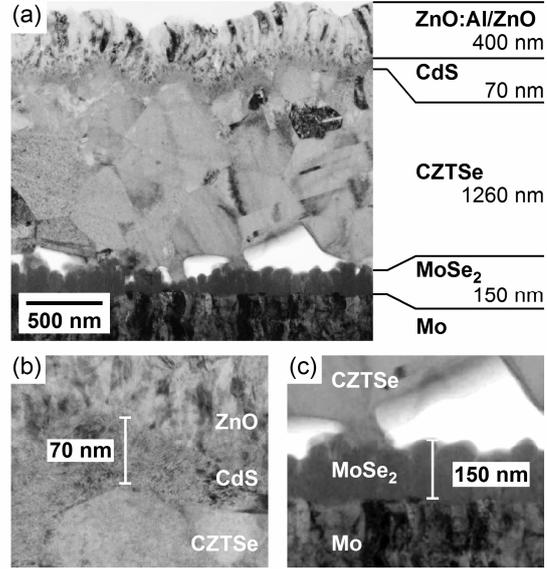

FIG. 1. Cross-sectional TEM images of the CZTSe solar cell fabricated in this study: (a) the whole solar-cell structure, (b) the enlarged image for the ZnO/CdS/CZTSe interface, and (c) the enlarged image for the CZTSe/Mo interface.

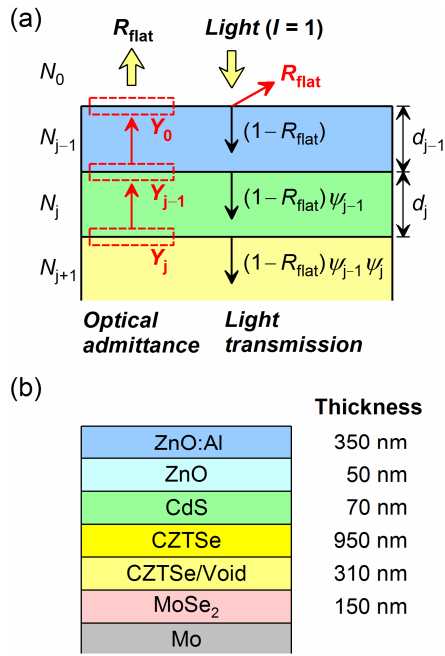

FIG. 2. (a) Calculation procedure of the optical admittance method and (b) optical model constructed for the CZTSe solar cell in Fig. 1.

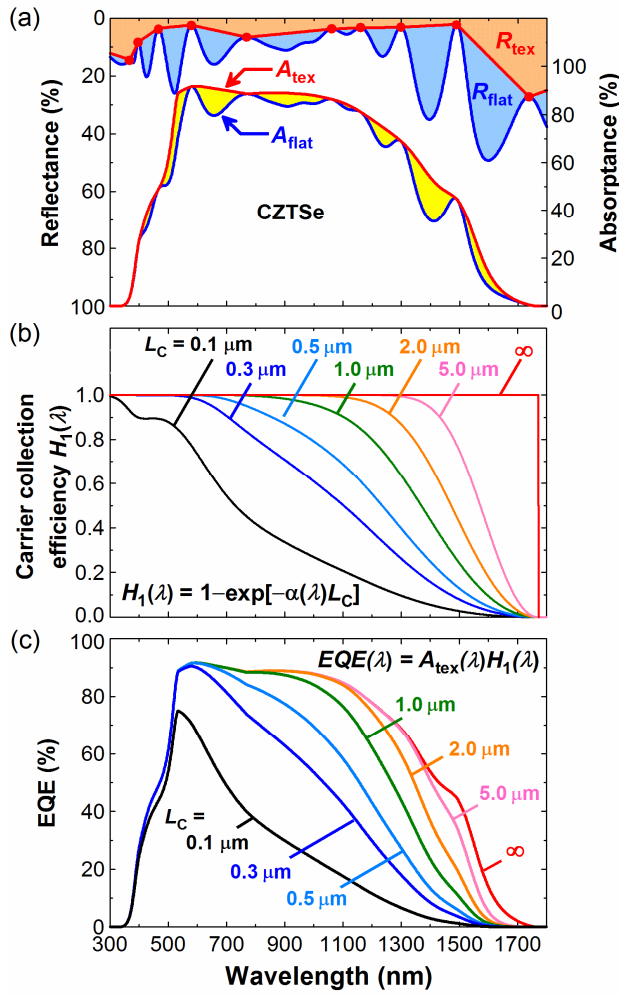

FIG. 3. Calculation procedure of the e-ARC method established in this study: (a) Reflectance spectra calculated assuming a flat structure ($R_{flat}$) and a textured structure ($R_{tex}$), together with the absorptance spectra of the CZTSe layer, $A_{flat}$ and $A_{tex}$, obtained using $R_{flat}$ and $R_{tex}$, respectively, (b) carrier collection efficiency $H_1(\lambda)$ calculated from the $\alpha(\lambda)$ of the CZTSe using different values of carrier collection length ($L_C$), and (c) EQE spectra of the CZTSe solar cell calculated based on the results of (a) and (b). This EQE simulation was performed using the optical model of Fig. 2(b).

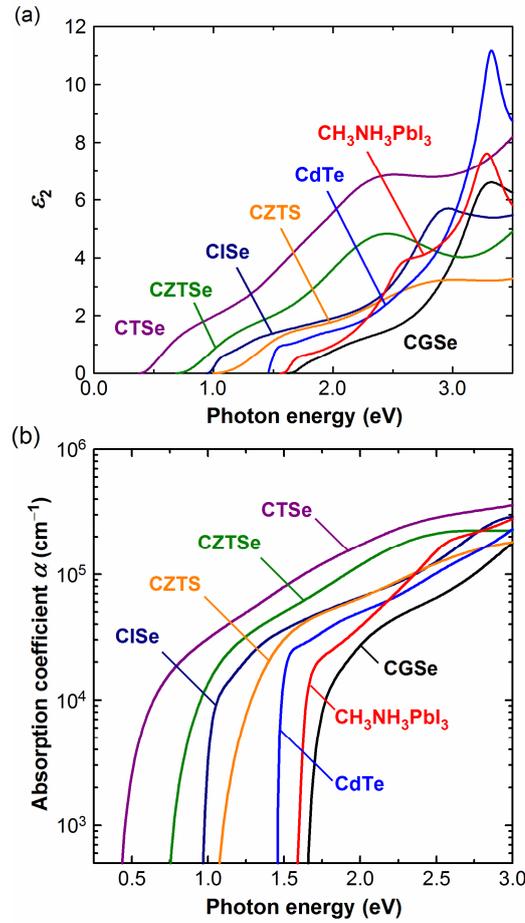

FIG. 4. (a) $\varepsilon_2$ spectra and (b) $\alpha$ spectra of various semiconductors applied for solar cells. In this figure, the reported optical data for CTSe,[61] CZTSe,[61] CISe,[66,67] CZTS,[68] CGSe,[66,67] CdTe,[69] and CH$_3$NH$_3$PbI$_3$[65] are shown.

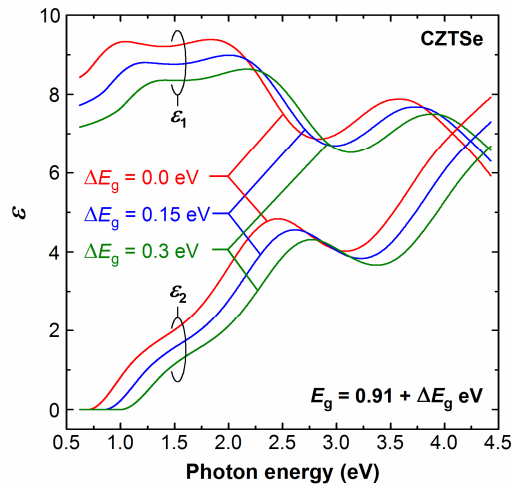

FIG. 5. Variation of the CZTSe dielectric function with the Cu content, calculated from Eq. (10) using different $\Delta E_g$ values. The dielectric function of $\Delta E_g = 0$ eV corresponds to that of the CZTSe in Fig. 4(a).

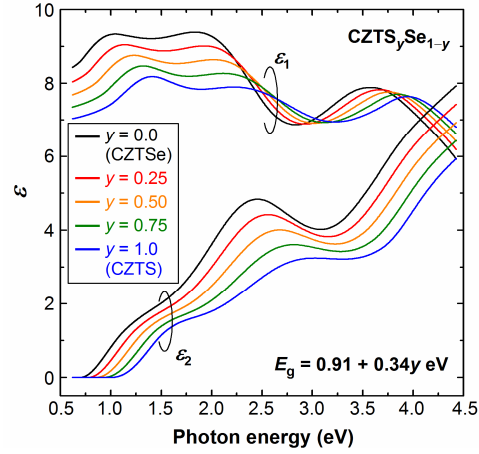

FIG. 6. Dielectric functions of CZTS$_y$Se$_{1-y}$ calculated by applying the energy shift model. The Cu content of the CZTS$_y$Se$_{1-y}$ is $x \sim 0.95$.

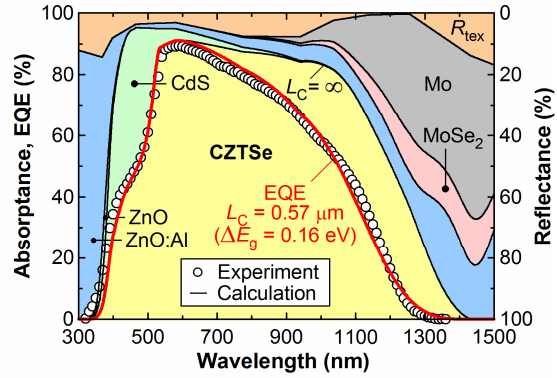

FIG. 7. Experimental EQE spectrum of the CZTSe solar cell consisting of ZnO:Al (350 nm)/ZnO (50 nm)/CdS (70 nm)/CZTSe (950 nm)/CZTSe-void (310 nm)/MoSe$_2$ (150 nm)/Mo (open circles), together with the fitted EQE result (red line) obtained from the e-ARC method assuming $L_C = 0.57$ $\mu$m and $\Delta E_g = 0.16$ eV. The absorptance spectra of the solar-cell component layers and $R_{tex}$ determined from the ARC method are shown by the black lines.

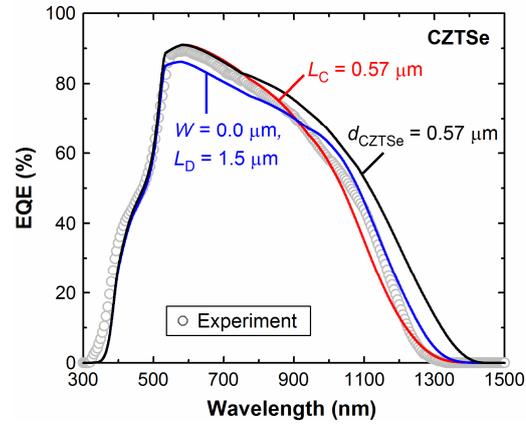

FIG. 8. Results of the EQE fitting analyses for the CZTSe solar cell using different $H(\lambda)$ functions and optical models. The red and blue lines represent the results of the EQE fitting analyses using $H_1(\lambda)$ [Eq. (5)] and $H_2(\lambda)$ [Eq. (7)], respectively, while the black line indicates the result when the total thickness of the CZTSe layer ($d_{\text{CZTSe}}$) is assumed to be 0.57 $\mu$m. The experimental data (gray open circles) and the calculation result (red line) correspond to those of Fig. 7.

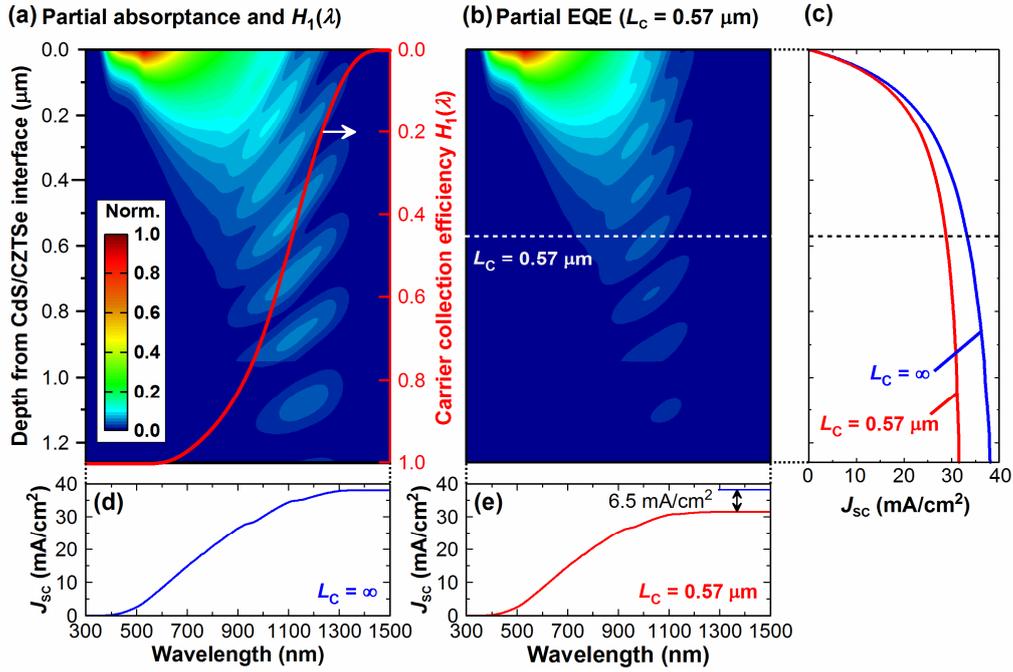

FIG. 9. (a) Partial $A$ of the CZTSe layer in the solar cell obtained at different depths from the CdS/CZTSe interface and wavelengths, together with $H_1(\lambda)$ obtained assuming $L_C = 0.57$ μm, (b) partial EQE of the CZTSe calculated assuming $L_C = 0.57$ μm, (c) integrated $J_{sc}$ for $d$, (d) integrated $J_{sc}$ for $\lambda$ when $L_C = \infty$ and (e) integrated $J_{sc}$ for $\lambda$ when $L_C = 0.57$ μm. In (a) and (b), the calculated values are normalized by the maximum values and the result of $d > 0.95$ μm corresponds to the CZTSe/void region in Fig. 2(b).

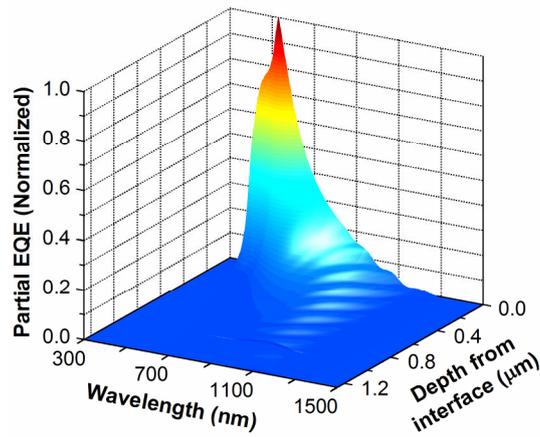

FIG. 10. Normalized partial EQE of the CZTSe solar cell. The partial EQE values correspond to those shown in Fig. 9(b).

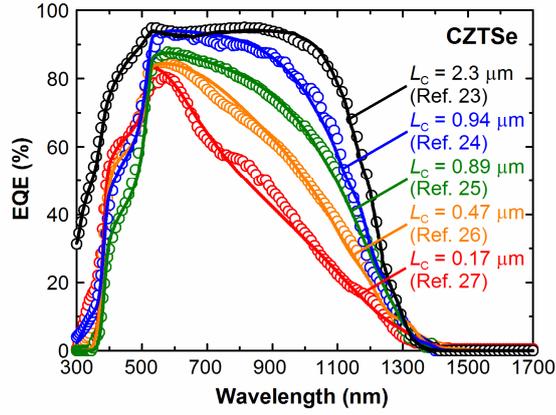

FIG. 11. Experimental EQE spectra of various CZTSe solar cells reported in Refs. 23-27 (open circles) and the results of the EQE fitting analyses performed for these solar cells using the e-ARC method (solid lines). The $L_C$ values obtained from the EQE analyses are also shown.

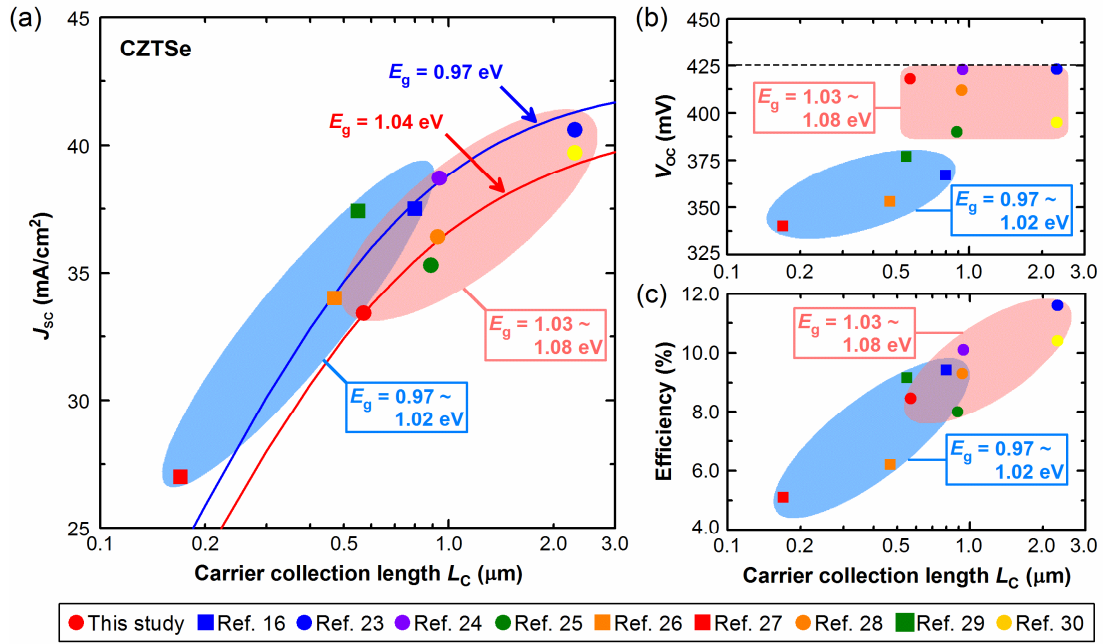

FIG. 12. (a) $J_{sc}$, (b) $V_{oc}$ and (c) efficiency of reported CZTSe solar cells (Refs. 16, 23-30) as a function of $L_C$ deduced from the EQE analyses using the e-ARC method. Based on the $E_g$ values obtained from the EQE analyses, the CZTSe active layers are categorized into two groups of $E_g = 0.97$–$1.02$ eV (closed squares) and $E_g = 1.03$–$1.08$ eV (closed circles). The solid lines show the variations of $J_{sc}$ with $L_C$, calculated assuming $E_g = 0.97$ eV (blue line) and $1.04$ eV (red line). For these simulations, a uniform CZTSe layer ($2.0$ $\mu$m thickness) without the void component is assumed in the optical model of Fig. 2(b).

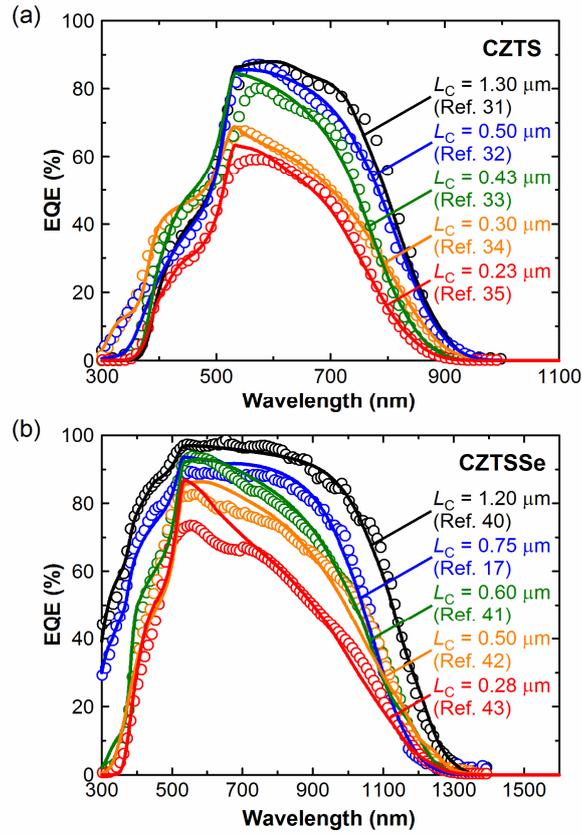

FIG. 13. Experimental EQE spectra of (a) CZTS solar cells reported in Refs. 31-35 and (b) CZTSSe solar cells reported in Refs. 17, 40-43 (open circles), together with the results of the EQE fitting analyses performed for these solar cells using the e-ARC method (solid lines). The $L_C$ values obtained from the EQE analyses are also shown.

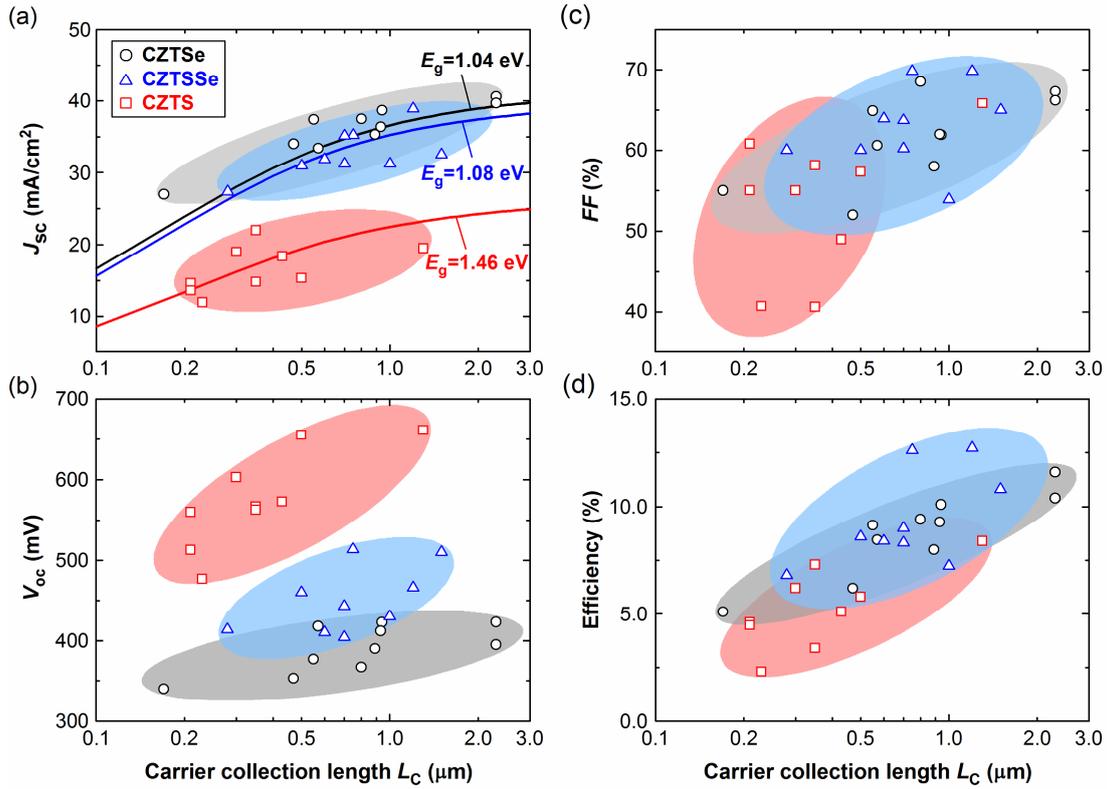

FIG. 14. (a) $J_{sc}$, (b) $V_{oc}$, (c) FF and (d) efficiency of reported CZTSe (Refs. 16, 23-30), CZTS (Refs. 31-39) and CZTSSe (Refs. 17, 40-47) solar cells as a function of $L_C$ deduced from the EQE analyses using the e-ARC method. The solid lines indicate the variations of $J_{sc}$ with $L_C$, calculated assuming the $E_g$ values of 1.04 eV (CZTSe), 1.08 eV (CZTSSe) and 1.46 eV (CZTS).

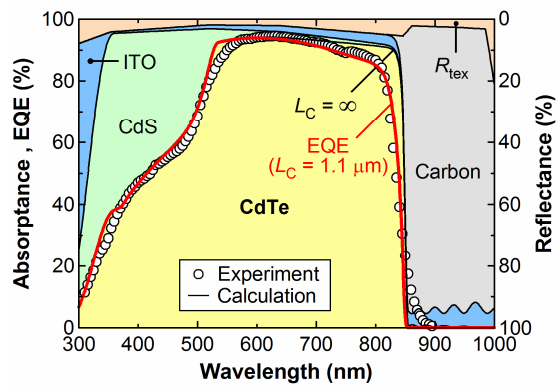

FIG. 15. Experimental EQE spectrum of a CdTe solar cell consisting of MgF$_2$/glass/ITO (200 nm)/CdS (50 nm)/CdTe (3.5 $\mu$m)/carbon electrode,[48] together with the fitted EQE result (red line) obtained from the e-ARC method assuming $L_C = 1.1$ $\mu$m. The $A$ spectra of the solar-cell component layers and $R_{tex}$ determined from the ARC method are also shown by the black lines.

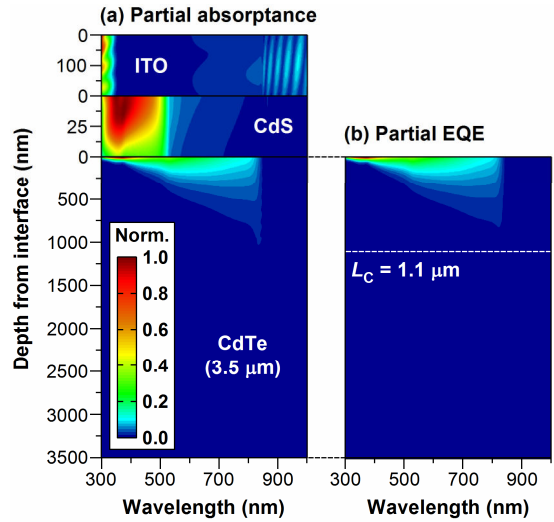

FIG. 16. (a) Partial $A$ for the ITO (200 nm)/CdS (50 nm)/CdTe (3.5 $\mu$m) and (b) partial EQE of the CdTe layer ($L_C$ = 1.1 $\mu$m) for the CdTe device in Fig. 15. The partial $A$ and EQE values are normalized by the maximum values in each layer.

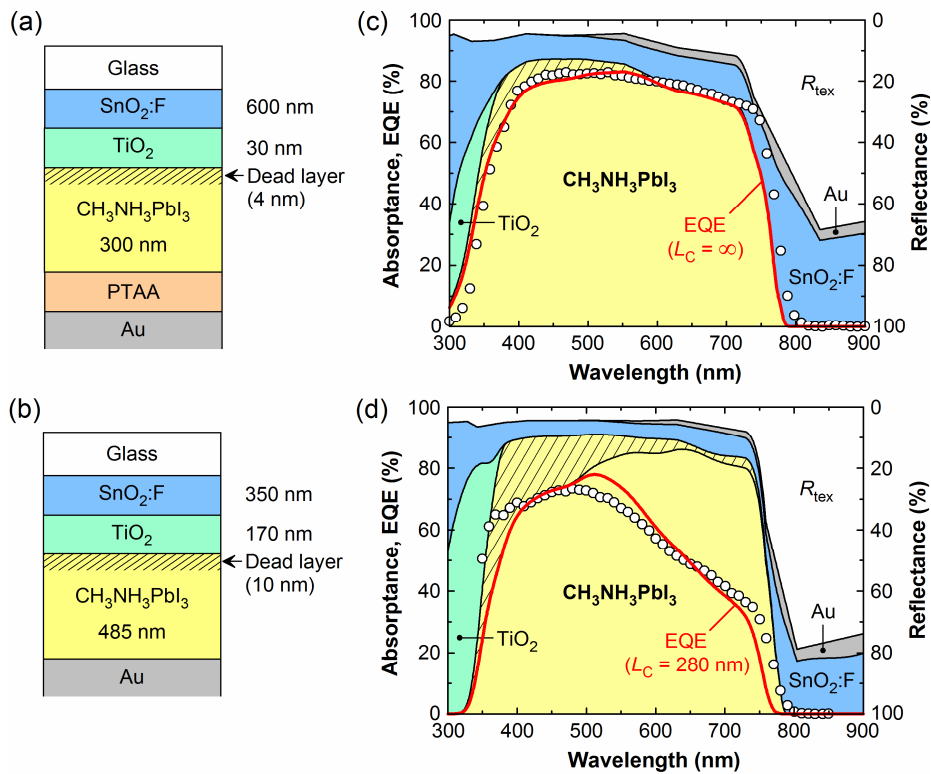

FIG. 17. Optical models for $CH_3NH_3PbI_3$ solar cells (a) with an HTL (Ref. 49) and (b) with no HTL (Ref. 50), and the corresponding EQE analysis results for the $CH_3NH_3PbI_3$ solar cells (c) with the HTL and (d) with no HTL. In (a) and (b), the values represent the layer thicknesses in the optical model, and the thickness of the PTAA layer in (a) is assumed to be zero. For the $CH_3NH_3PbI_3$ layers, the presence of optical dead layers near the $TiO_2$ interface is assumed.

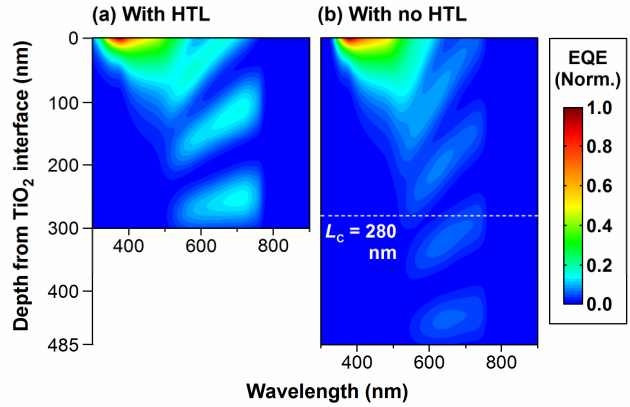

FIG. 18. Partial EQE obtained at different depths from the $TiO_2$/$CH_3NH_3PbI_3$ interface in the $CH_3NH_3PbI_3$ solar cells (a) with the HTL and (b) with no HTL. These partial EQE values correspond to the EQE spectra shown as the red lines in Fig. 17. The partial EQE values are normalized by the maximum values in each solar cell. In (b), the dotted line indicates the position of $L_C$ = 280 nm.

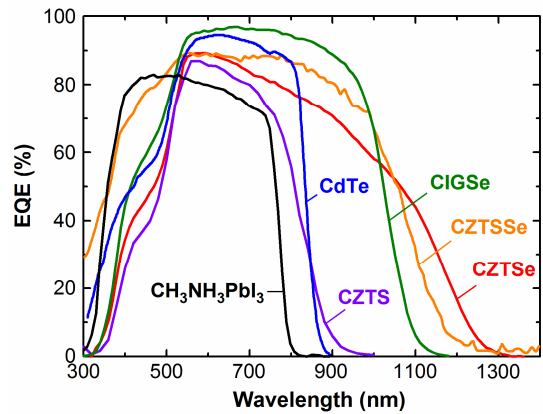

FIG. 19. Experimental EQE spectra for various solar cells: CZTSe (Fig. 7), CZTSSe [Ref. 17 in Fig. 13(b)], CIGSe (Ref. 19), CZTS [Ref. 31 in Fig. 13(a)], CdTe (Ref. 48 in Fig. 15) and $CH_3NH_3PbI_3$ [Ref. 49 in Fig. 17(c)].

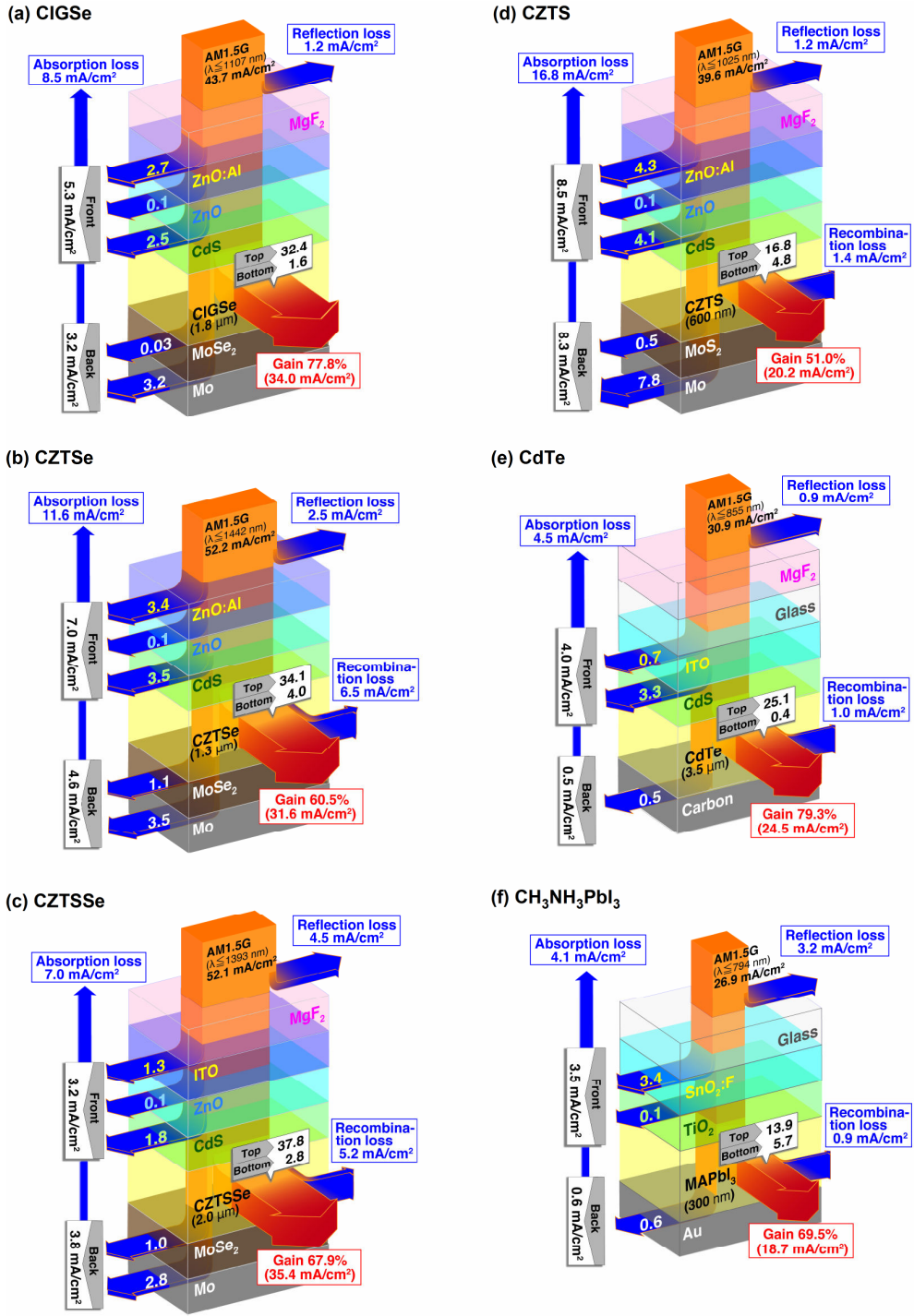

FIG. 20. Optical and recombination losses deduced from the EQE analyses performed for the spectra of Fig. 19: (a) CIGSe, (b) CZTSe, (c) CZTSSe, (d) CZTS, (e) CdTe, and (f) $CH_3NH_3PbI_3$ (MAPbI$_3$). The numerical values represent the corresponding current densities in units of mA/cm$^2$. The maximum $J_{sc}$ value attainable under AM1.5G condition and the $J_{sc}$ contributions obtained assuming two sublayers (top and bottom layers) with an equal thickness are also indicated. The optical gain represents the ratio of output $J_{sc}$ divided by the maximum-attainable $J_{sc}$ value. The result of the CIGSe is adopted from Ref. 19.

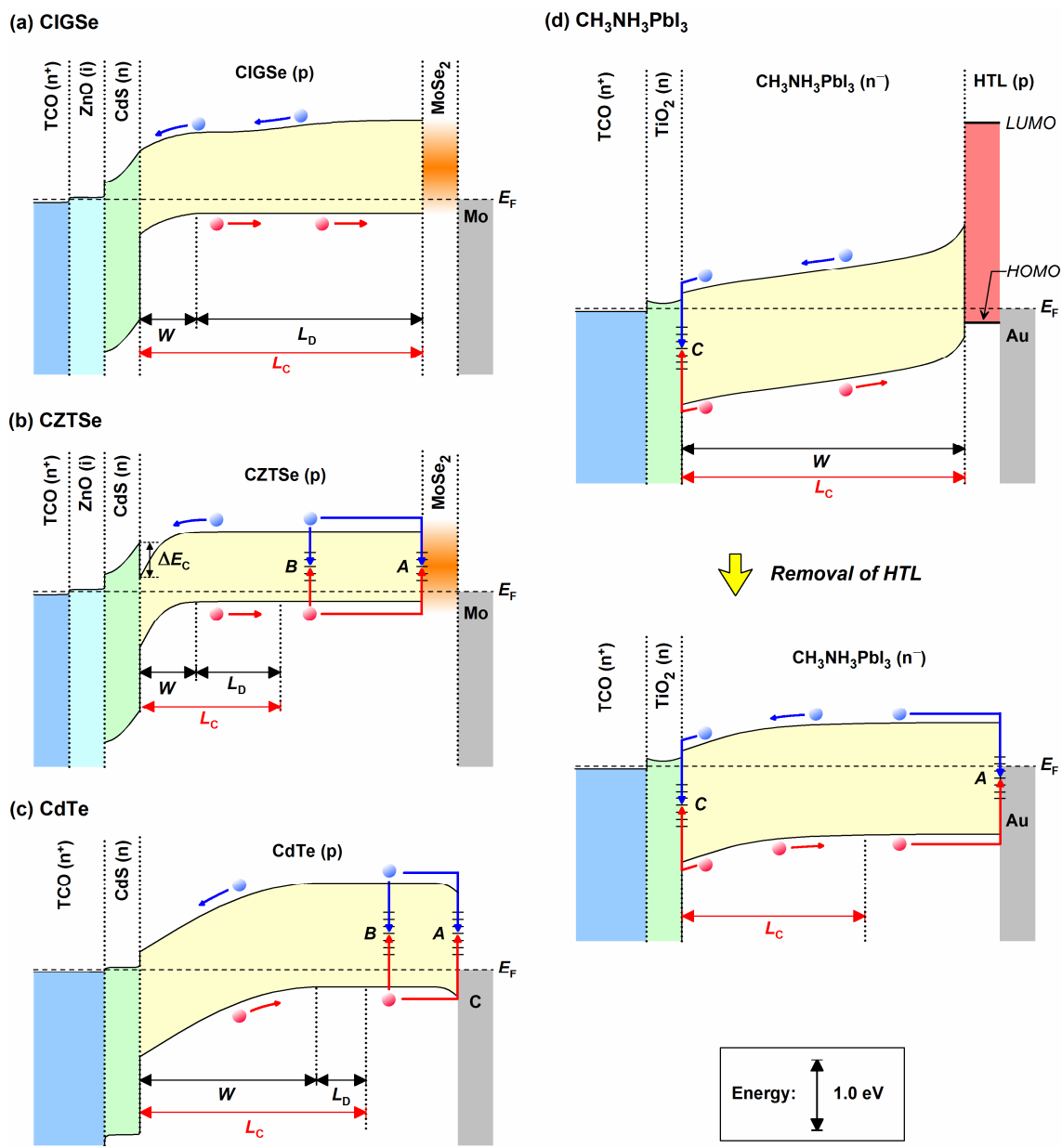

FIG. 21. Band diagrams for various solar cells of (a) CIGSe, (b) CZTSe, (c) CdTe, and (d) $CH_3NH_3PbI_3$ with an HTL (spiro-OMeTAD) and with no HTL. In this figure, *A*, *B* and *C* represent the recombination processes in the rear interface, bulk and front interface regions, respectively.

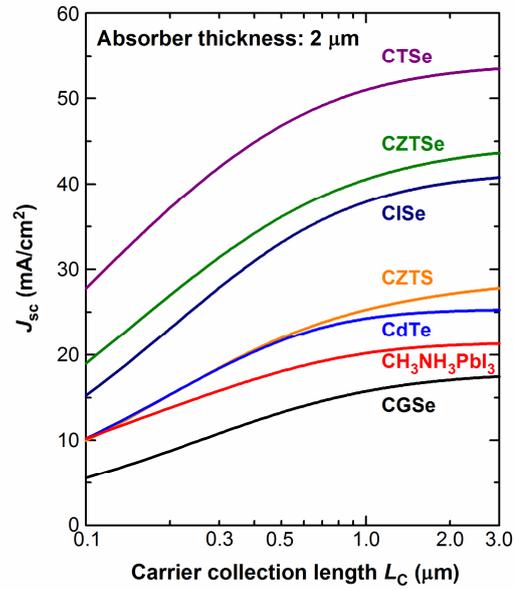

FIG. 22. Variation of $J_{sc}$ with $L_C$ for various solar cells calculated from the e-ARC method. For all the solar cells, an identical solar cell structure consisting of the MgF$_2$ (130 nm)/ZnO:Al (360 nm)/ZnO (50 nm)/CdS (45 nm)/absorber (2.0 $\mu$m)/Mo is assumed, except for CdTe and CH$_3$NH$_3$PbI$_3$ solar cells. For the CdTe and CH$_3$NH$_3$PbI$_3$ solar cells, the structures in Fig. 15 (CdTe) and Fig. 17(a) (CH$_3$NH$_3$PbI$_3$) were adopted, but the CdTe and CH$_3$NH$_3$PbI$_3$ thicknesses were assumed to be 2.0 $\mu$m.